\documentclass[12pt]{article}%
\usepackage{mathpazo}%
\usepackage{pifont}%
\usepackage[fleqn]{amsmath} %
\usepackage{amssymb,latexsym} %
\usepackage{tocloft,float}%
\usepackage[dvipsnames,svgnames,x11names]{xcolor}
\usepackage{tikz}%
\usepackage[colorlinks=true,linkcolor=DarkRed!,
  urlcolor=DarkRed!,citecolor=DarkRed!]{hyperref}
\newcommand{\ncom}{\newcommand}%
\newcommand{\rcom}{\renewcommand}%
\rcom{\mathindent}{\parindent}%
\rcom{\baselinestretch}{1.20}%
\newenvironment{refs}%
{\begin{list}{}%
 {\setlength{\labelwidth}{0pt} %
  \setlength{\leftmargin}{\parindent} %
  \setlength{\itemindent}{-1.0\parindent}%
  \setlength{\itemsep}{-0.5ex}%
   } }%
{\end{list}}
\rcom{\contentsname}{\normalsize Table of Contents}%
\rcom{\cftsecfont}{\small }%
\rcom{\cftsubsecfont}{\small }%
\rcom{\cftdotsep}{6} %
\rcom{\cftsecleader}{\cftdotfill{\cftdotsep}}%
\rcom{\cftsecpagefont}{\normalfont\small } %
\rcom{\cftaftertoctitleskip}{-15pt} %
\rcom{\cftsetrmarg}{2em} %
\rcom{\cftsetpnumwidth}{1em} %
\ncom{\mrm}{\mathrm}%
\ncom{\tsf}{\textsf}%
\ncom{\trm}{\textrm} %
\ncom{\tbf}{\textbf} %
\ncom{\sfS}{\tsf{S}} %
\ncom{\sfM}{\tsf{M}} %
\ncom{\sfe}{\tsf{e}}%
\ncom{\R}{\ensmath{\mathbb{R}}}%
\ncom{\N}{\ensmath{\mathbb{N}}}%
\ncom{\C}{\ensmath{\mathbb{C}}}%
\ncom{\rmi}{\trm{i}}%
\ncom{\mm}{\textrm{m}}%
\ncom{\la}{\langle}%
\ncom{\ra}{\rangle}%
\ncom{\calH}{\mathcal{H}}%
\ncom{\calI}{\mathcal{I}}%
\ncom{\calS}{\mathcal{S}}%
\ncom{\calE}{\mathcal{E}}%
\ncom{\calB}{\mathcal{B}}%
\rcom{\to}{\rightarrow}%
\ncom{\bm}[1]{\mbox{\boldmath{$#1$}}}%
\ncom{\uniek}{\bm{\exists!}}%
\ncom{\ensmath}{\ensuremath}%
\ncom{\punt}{\protect\hspace*{-1em}\tbf{.}~~}%
\ncom{\bfbox}{\ensmath{\bm{\Box}}}%
\ncom{\zop}{\mbox{$|\!\uparrow\ra$}} %
\ncom{\zneer}{\mbox{$|\!\downarrow\ra$}} %
\ncom{\exmeer}{\\[1ex]}%
\ncom{\emmeer}{\\[1em]}%
\ncom{\half}{\mbox{\small $\frac{1}{2}$}}%
\ncom{\one}{\emph{1}}%
\ncom{\zero}{\emph{0}}%
\ncom{\fn}{\footnote}%
\ncom{\terug}{\protect\hspace*{-1ex}}%
\ncom{\beq}{\begin{equation}}%
\ncom{\enq}{\end{equation}}%
\ncom{\donkrood}[1]{\textcolor{DarkRed}{#1}}%
\ncom{\display}[1]{\\[0.8em]\indent{\parbox[t]{0.95\textwidth}{#1}}\\[0.8em]}%
\begin{document}%
\raggedbottom%
\thispagestyle{empty}
\noindent%
{\small Will appear as a Chapter in: \emph{Non-Reflexive Logics, Non-Individuals and the Philosophy of Quantum Mechanics: Essays in honor of the philosophy of D\'{e}cio Krause}, Springer-Verlag, 2023.}\\*[3em]
{\LARGE \textbf{\donkrood{Six Measurement Problems of Quantum Mechanics}}}\\[2em]
{\Large \emph{\textbf{F.A.\ Muller$\,$}}}\fn{Erasmus School of Philosophy, Erasmus University Rotterdam, and 
Faculty of Science, Utrecht University, The Netherlands;
e-mail: {\tt f.a.muller@uu.nl}.}\exmeer
Number of words: c.a.\ 12,000 (including references and footnotes).\exmeer
\emph{Corrected 10 June 2023}
\begin{quote}
{\small \textbf{Summary.}~~The notorious `measurement problem' has been roving around quantum mechanics for nearly a century since its inception, and has given rise to a variety of `interpretations' of quantum mechanics, which are meant to evade it. We argue that no less than 
\emph{six} problems need to be distinguished, and that several 
of them classify as \emph{different types} of problems. One of
them is what traditionally is called `the measurement problem'. Another of them has nothing to do with
measurements but is a profound metaphysical problem.
We also analyse critically T.\ Maudlin's (1995)
well-known statement of `three measurements problems',
and the clash of the views of H.~Brown (1986) and H.~Stein (1997) on one of the six meansurement problems. Finally,
we summarise a solution to one measurement problem which
has been largely ignored but tatictly if not explicitly acknowledged.\mbox{}}
\tableofcontents
\end{quote}
\thispagestyle{empty}
\clearpage
\setcounter{page}{1}
\section{{\punt}Exordium}
The other day, I wondered: \emph{who} discovered `the measurement 
problem' of Quantum Mechanics (QM), and who coined it?
If the problem is that superpositions carry over from
`microscopic' physical systems (molecules, atoms, 
particles) to 
`macroscopic' physical systems, then Einstein discovered 
it, when he wrote in 1935 to Schr\"{o}dinger about a bom in an 
exploded and non-exploded state.\fn{Fine (1986, Ch.$\,$5).} As is well-known, 
Schr\"{o}dinger went public with this, but not before replacing the 
bomb with a cat in a superposition of states of the cat being 
alive and being death, making the cat neither dead nor alive. 
Recently Rovelli prefered to 
consider a cat in a friendlier superposition, of being wide awake and being purring asleep, and Norson a cat with a fat belly having drunk milk and a cat with an empty belly
having drunk no milk.\fn{Rovelli (2021, Ch.$\,$2), Norsen (2017, p.$\,$73).}

If `the measurement problem' is however that when 
we describe the \emph{measurement interaction} unitarily and end up 
with a measurement apparatus indicating no definite
measurement outcome, then Von~Neumann discovered the problem.
Von~Neumann inaugurated quantum-measurement theory in 
his magisterial \emph{Mathematische Grundlagen der 
Quantenmechanik} (1932), and introduced his notorious 
Projection Postulate in order to end up with a single definite 
measurement outcome; this is \emph{stricto sensu} a solution to `the measurement problem'.\fn{Neumann (1932); for a thorougly updated version, including a deposit story of results from mathematical physics about quantum theory achieved after 1932, see Landsman (2017).} Did Einstein obtain the explosive idea of the bomb from Von~Neumann's measurement theory, both members of the Princeton Institute for Advanced Studies at the time, and Einstein being familiar with Von~Neumann's \emph{Grundlagen}? 

These questions have, I must sadly report, no definite
answers. A quick survey in historical writings on QM came up 
empty, and posing the question on the e-mail list of 
the hopos community 
has taught me there are no definite answers to these 
questions. A few pertinent remarks merit mentioning,
historical underdeterminacy notwithstanding.

One remark is that since Einstein's and Von~Neumann's
considerations saw the light of day earlier than Schr\"{o}dinger's, the Columbus Price for Landmark 
Discoveries does not go to this Austrian pussycat.
Another remark is that the dawn of talk of `the measurement 
problem' lies in the early 1960s, with Wigner (1961, 
1963), which makes Wigner the undisputed prime 
candidate for `Eugene Paul (\emph{n\'{e}} Jen\"{o} P\'{a}l) the Baptist' of
`the problem of measurement' of QM; in both mentioned papers, 
Wigner expounds `the measurement problem' with crystal clarity
--- and then unhesingtately solves it by invoking 
the Projection Postulate. All expositions of QM, 
whether aimed at working physicists, mathematicians or students, then and even now, have included and do include the Projection Postulate, as being part and parcel of standard QM, whether their authors accept, doubt or reject it; and then there is no `measurement problem'. Unless `the measurement problem' is how to get rid of the Projection Postulate.

Enough history. Before we proceed to state no less than
six measurement problems (in Sections~\ref{SectRPMO},
\ref{SectRPCW}, \ref{SectMIP} and \ref{SectMEP}), we need to state some of the 
postulates of standard QM precisely, if only for the sake of 
reference. We compare the very first problem,
the Reality Problem of Measurement Outcomes, 
critically with Maudlin's
well-known exposition of `three measurement problems'
(Section~\ref{SectComp}).
`Insolubility Theorems' and the impossibility of
describing measurement interactions unitarily 
are the topic of Section~\ref{SectInsolu}, as well as
Brown's (1986) and Stein's (1997) clashing take on these theorems. Section~\ref{SectMEP} includes a summary of an explication of the concept of measurement, which is the most ignored measurement problem of the six.
We recaputilate and draw some conclusions at the end (Section~\ref{SectRecap}).
\section{{\punt}Some Postulates of Standard Quantum Mechanics}
\label{SectPostQM}
The primitive physical concepts of the vocabulary of standard QM, i.e.\ the ones without
definitions, are: \emph{physical system,
subsystem, measurement cq.\ measurement apparatus, property, 
space, time, state, probability} and \emph{physical magnitude} 
(Dirac (1928) called physical magnitudes positivistically `observables', a 
terminology that has stuck; Von~Neumann spoke of `physical
quantities', which terminology is also in use). 
Sometimes one speaks of `physical variables', which is troublesome for certain reasons and not troublesome for different reasons (we park them).
A \emph{subsystem} of a physical system is a mereological
part of it; the relation `is a subsystem of' coincides with
the part-whole relation in mereology and is governed by its axioms. 

The first two postulates tell us what the mathematical representatives are of the physical states and physical magnitudes.\fn{A square (\bfbox) marks a postulate of standard QM;
a black box ($\blacksquare$) marks a principle worth
considering; these are both written in \emph{italics}. A dark red triangle (\donkrood{\ding{115}}) signals a problem: there are six of them: I--VI.}
\display{\tbf{\bfbox~~Pure State Postulate.} \emph{Pure physical states of a physical system $\sfS$ are represented by unit-vectors in some Hilbert-space ($\calH$), up to a multiplicative complex constant of modulus equal to~$1$, a `phase factor' $\mrm{e}^{\mrm{i}\theta}$. Whenever physical system $\sfS$ consists of $N$ subsystems, its Hilbert-space $\calH$ consists of the $N$-fold direct-product Hilbert-space of the $N$ subsystems}:
$\,\calH\,=\,\calH_{1}\otimes\calH_{2}\otimes\;\ldots\;
\otimes\calH_{N}$.\emmeer
\tbf{\bfbox~~Magnitude Postulate.}
\emph{Every physical magnitude pertaining to physical 
system $\sfS$ is represented by
some self-adjoint operator $A$ (up to a real multiplicative 
constant) that acts on the Hilbert-space $\calH$ of $\sfS$.}}
Needless to say that the restriction to self-adjoint operators
can be loosened to e.g.\ normal operators
(they commute with their adjoint) or to positive operators 
--- one can prove a spectral theorem for normal operators, but not for positive operators, which makes positive operators  mathematically minacious. 
Such loosenings will have no bearing on the problems (and their possible solutions) treated in this paper.
Notice that the troublesome converse is \emph{not} part
of the Magnitude Postulate: not \emph{every} self-adjoint operator needs to represent a physical magnitude.\fn{Recall Wigner's famous question: which unmeasurable physical magnitude represents the self-adjoint operator $P+Q\,$?} 

The following postulate connects magnitude operators 
to measurements.
\display{\tbf{\bfbox~~Spectrum Postulate.}
\emph{All measurement-outcomes of a physical magnitude belong 
to the spectrum of the representing self-adjoint operator.}}

Spectral Theorems inform us that every self-adjoint (and every  normal) operator $A$ has a unique spectral family of projectors, \mbox{$\,P^{A}(\Delta):\calH\to\calH$}, where \mbox{$\,\Delta\in\calB(\R)\,$} is a Borel subset of~$\R$. We denote by $\calH(A,\Delta)$ the subspace of $\calH$ onto which $P^{A}(\Delta)$ projects, and by $\calH(A,a)$ the subspace of $\calH$ onto which $P^{A}(\{a\})$ projects. Further, we represent a \emph{determinate physical property} by an ordered pair $\la A,a\ra$, where $\,a\in\R$ is a member of the spectrum of $A$; to attribute $\la A,a\ra$ to a physical system is the same as: assigning value $a$ to physical magnitude (represented by operator) $A$. In consonance with current terminology in metaphysics, we shall call physical magnitude $A$ that pertains to physical system $\sfS$ a \emph{determinable physical property} of~$\sfS$.

A postulate that has become known under the misnomer `the eigenstate-eigenvalue link' provides a criterion for the
ascription of determinate properties to physical systems
depending on their state:
\display{\tbf{\bfbox~~Eigenlink.}
\emph{A physical system $\sfS$ 
having pure physical state $\,|\psi\ra\in\calH\,$ has determinate physical property $\la A,a\ra\;$ iff  
$\;|\psi\ra$ lies in the eigen-subspace of $A$ that belongs to $a$, that is, $\,|\psi\ra\in\calH(A,a)$.}}

This Eigenlink is limited to operators with a discrete spectrum. A generalistion to all types of spectra, discrete, continuous and combinations thereof, is possible, provided one is prepared to attribute `vague' properties, mathematically represented by $\la A,\Delta\ra$:
\display{\tbf{\bfbox~~Generalised Eigenlink.}
\emph{A physical system $\sfS$ 
having pure physical state $\,|\psi\ra\in\calH\,$ has determinate  physical property $\la A,\Delta\ra\,$, where $\Delta\subset\R$ is an interval,  
iff $\,|\psi\ra$ lies in the eigen-subspace of $A$,
that is, $\,|\psi\ra\in\calH(A,\Delta)$.}}

The Eigenlink is the special case of the Generalised Eigenlink when the spectrum is discerte and $\Delta$ is the singleton-set of some single eigenvalue: $\Delta=\{a_{j}\}$.

The Generalised Eigenlink is how Von~Neumann put it in \emph{Gundlagen}, as a corollary of representing properties by  projectors (1932, item $(\beta)$, p.$\,$253); it is a straightforward generalisation of the Eigenlink. But one may frown about a property that is mathematically represented by $\la A,\Delta\ra$, due to $\Delta$ being a \emph{subset} of $\R$ and generically containing non-denumerably many values of $A$. From the attribution of $\la A,\Delta\ra$ to physical system $\sfS$, we are \emph{not} supposed to infer, absurdly,  that $\sfS$ then jointly possesses property $\la A,a\ra$ for every $\,a\in\Delta$ that lies in spectrum of~$A$. Some have argued this is yet another quantum-mechanical novelty, not entirely unfamiliar to metaphysicists: a \emph{determinate vague} property, with sharp boundaries, so not vague in the standard sense.\fn{See e.g.\ French and Krause (1995), and Bush \emph{et al.}\ (1990, p.$\,$127), who talk about ``vague objectification''.} If one rejects the idea of a vague property, perhaps because it is unfathomable, then one can remain aboard with the Eigenlink and its sharp determinate properties, and throw the Generalised Eigenlink overboard.

Finally, a postulate that is not a postulate of standard QM,
but is a postulate of nearly every other intepretations of QM, is the following one. 
\display{$\blacksquare$~~\tbf{Universal Dynamics Postulate.}
\emph{Time is represented by the real continuum $(\R)$.
For every physical system $\sfS$, there is some 
connected continuous Lie-group
of unitary operators $\,t\mapsto U(t)\,$ acting on $\calH$
such that the state at time $t$ is 
$\,U(t)|\psi(0)\ra=|\psi(t)\ra\,$ when $|\psi(0)\ra$
is the state at time $t=0$.}}
This Lie-group of \emph{time-translations} 
is the solution of the Schr\"{o}dinger 
equation for the self-adjoint Hamiltonian $H$, the operator that represents energy. Operator $H$ characterises the 
physical system and determines how the state of the system changes over time via the Stone-Von~Neumann Theorem, which  theorem associates such a Lie-group uniquely with every self-adjoint operator by means of the equation: 
\mbox{$\,U(t)=\exp[-\rmi Ht/\hslash]$}. The ~\tbf{\bfbox~Dynamics Postulate} of standard QM says the same as the universal one, but \emph{conditional} on that no measurements are 
performed (\emph{lege infra}). 

For the sake of clarity, \emph{standard} QM is the theory defined by the State, Magnitude, Spectrum, Probability, Dynamics, Projection, and Symmetry Postulate, and the  Eigenlink.\fn{Some of the postulates just mentioned have not been stated yet; they will be stated below, when they are needed.}
\section{{\punt}The Reality Problem of Measurement Outcomes}
\label{SectRPMO}
What is generally known as `the measurement problem',
we shall call `the Reality Problem of Measurement Outcomes'.
We state it as a logical incompatibility of
five propositions, taking for granted the relevant parts of 
mathematics relied on. Before stating the problem,
we need to express one more proposition:
\display{$\blacksquare$~~\tbf{Single Measurement Outcome Principle (SMOP).} \emph{Every performed measurement has a single outcome, 
provided the pieces of measurement apparatus involved
do not malfunction.}}

SMOP seems very much a universal empirical regularity: measurements obtaind by properly functioning pieces of measurement apparatus always have \emph{single outcomes}. In cases where there is \emph{no outcome}, some involved piece of equipment malfunctioned. In case there is \emph{more than one} outcome $\ldots$ But that seems impossible. How \emph{can} a LED or LCD display show more than one number? How \emph{can} a pointer at any moment of time indicate more than one mark on a scale? Are \emph{such} measurement events not simply physically impossible? Do we really need a \emph{principle} (SMOP) to rule out what seem to be physically impossible? 

SMOP certainly seems a universal empirical regularity
fully supported by the practice of performing measurements. 
But we need to state it nonetheless to make the proof below devoid of any logical gaps; and furthermore, we know that there are interpretations of QM that \emph{reject} SMOP, e.g.\ the Everett and the Many Worlds Interpretation.

We arrive at the first problem.\fn{Bush \emph{et al.}\
(1996, p.$\,$91 ff.) call it, curiously, 
``the objectification problem''.} This problem is what we call a \emph{polylemma}: to reject at least one of any number
of propositions because they are shown to be jointly inconsistent.
\display{\donkrood{\ding{115}}~\tbf{I. Reality Problem of Measurement Outcomes.} Granted the relevant background mathematics, and given the State, Magnitude and Spectrum Postulate; then the Universal Dynamics Postulate, the Property Revealing Condition (\emph{lege infra}), the Single Measurement Outome Principle (SMOP), and the Eigenlink are 
jointly inconsistent. \label{RPMO}}

\emph{Proof.} Consider
the famous Stern-Gerlach experiment, where one performs 
measurements of the spin of a charged particle after it has 
passed the inhomogenous magnetic field of a DuBois magnet.
We are going to apply the mentioned postulates and principles to this experiment, which yields a QM-model of this experiment, and show how they clash logically.

We have an electron ($\sfe$) and a piece of measurement apparatus ($\sfM$), with Hilbert-spaces $\,\calH_{\sfe}=\C^{2}\,$ and $\,\calH_{\sfM}=\C^{3}$, respectively,
and Hilbert-space $\,\calH_{\sfe}\otimes\calH_{\sfM}=\C^{6}\,$ for the composite system ($\sfe\sqcup\sfM$,
State Postulate). Pauli-matrix
$\sigma_{z}$, an operator acting on $\C^{2}$, represents $z$-spin (Magnitude Postulate), which has two orthogonal eigenvectors 
in $\C^{2}$. The measurement-magnitude
(pointer-magnitude, display-magnitude) we represent
by operator
$M$, which acts in $\C^{3}$; $M$ has by definition
three orthogonal eigenstates, with the following associated determinate properties (Eigenlink):
\beq\begin{array}{lll}
\zop &:~~\sfe~~\trm{has determinate property}&\la\sigma_{z},\uparrow\ra\,;\exmeer
\zneer &:~~\sfe~~\trm{has determinate property}&\la\sigma_{z},\downarrow\ra\,;\exmeer
|+\ra &:~~\sfM~~\trm{has determinate property}&\la M,m_{\uparrow}\ra\,;\exmeer
|-\ra &:~~\sfM~~\trm{has determinate property}&\la M,m_{\downarrow}\ra\,;\exmeer
|m_{0}\ra &:~~\sfM~~\trm{has determinate property}&\la M,m_{0}\ra\,.
\end{array}\enq
In state $|m_{0}\ra$, the measurement device $\sfM$ 
has been turned on
and is ready to measure; $\sfM$ is prepared in this state
before the measurement begins. Both $\sigma_{z}$
and $M$ are self-adjoint operators.\fn{The numerical 
eigenvalues that
$\uparrow$ and $\downarrow$ symbolise are $+\hslash/2$
and $-\hslash/2$, respectively. The values $m_{\uparrow}$, 
$m_{\downarrow}$ and $m_{0}$ can be chosen arbitrarily, provided they are different, e.g.\
$m_{0}=5$, $\,m_{1}=m_{\uparrow}=+1$, 
$m_{2}=\,m_{\downarrow}=-1$; then $\,M|m_{j}\ra=m_{j}|m_{j}\ra\,$, for $\,j\in\{0,1,2\}$.\label{fnmj}}

We are going to measure $z$-spin of the electron, 
a process that takes $\tau$ seconds, say. THe interaction
between $\sfe$ and $\sfM$, codified by the
Hamiltonian, is supposed to be a measurement
interaction; it determines $\,t\mapsto U_{\mm}$ (Universal Dynamics Postulate). 
But what is a measurement interaction, an interaction between a physical system that is being measured and one that is doing the measuring? 

One straightfoward necessary condition reads
that if the measured system possesses some determinate property
that is measured, then the measured apparatus must reveal 
it (an instance of the Property Revealing Condition, see
remark \tbf{3$\mathbf{^\circ}$} below):
\beq 
U_{\mm}(\tau)\big(\zop\otimes|m_{0}\ra\big)\,=\,\zop\otimes|+\ra\quad\trm{and}\quad
U_{\mm}(\tau)\big(\zneer\otimes|m_{0}\ra\big)\,=\,\zneer\otimes|-\ra\,.
\label{Um}
\enq

Both initial and final states in \eqref{Um} 
are eigenvectors of {$\,\sigma_{z}\otimes M$}.
On the basis of the Eigenlink, we then can
assign the correct determinate physical properties to
$\sfe$ and to $\sfM$. The Spectrum Postulate
says that upon measurement of $z$-spin, we can only
find, as measurement outcomes, the two eigenvalues of 
$\sigma_{z}$: $\uparrow$ and $\downarrow$. 
Semantic convention has it that physical magnitude 
$\sigma_{z}$ having
value $\uparrow$ or $\downarrow$ is the same as saying that $\sfe$ has determinate property {$\,\la\sigma_{z},\uparrow\ra$}
or {$\,\la\sigma_{z},\downarrow\ra$}, respectively,
and that $\sfM$ indicating outcome $m_{\uparrow}$ or $m_{\downarrow}$ is the same as $\sfM$ possessing determiate property {$\la M,m_{\uparrow}\ra$} or {$\la M,m_{\downarrow}\ra$}, respectively. 

Suppose that initially, at time $t=0$, the composite system is in the following state:
\beq
|\psi(0)\ra\,=\,\big(\alpha\zop+\beta\zneer\big)\otimes|m_{0}\ra\,,\enq
with $\,\alpha,\beta\in\C\,$ being both non-zero, and 
$\,|\alpha|^{2}+|\beta|^{2}=1$.
At time $\,t=\tau$, the post-measurement state of the composite system 
is, due to the Universal Dynamics Postulate
and requirement~\eqref{Um}:
\beq\begin{array}{ll}
|\psi(\tau)\ra &=\;U_{\mm}(\tau)|\psi(0)\ra \exmeer
&=\;U_{\mm}(\tau)\,\alpha\zop\otimes|m_{0}\ra+
U_{\mm}(\tau)\,\beta\zneer\otimes|m_{0}\ra\exmeer
&=\;\alpha\zop\otimes|+\ra\;+\;\beta\zneer\otimes|-\ra\,.
\end{array}\label{psitau}\enq
By the Eigenlink, at time $t=\tau$, since the state of the composite system is not an eigenvector of operator 
{$\,\sigma_{z}\otimes M$}, $\sfe$ does neither have the determinate
$z$-spin property up, {$\la\sigma_{z},\uparrow\ra$},
nor down {$\la\sigma_{z},\downarrow\ra$}, and 
$\sfM$ does neither have the determinate property {$\la M,m_{\uparrow}\ra$}
nor {$\la M,m_{\downarrow}\ra$}, and therefore does not indicate
an outcome. This contradicts SMOP.  ~\emph{Q.e.d.}
\emmeer
We end this Section with a number of systematic remarks.

\tbf{1$\mathbf{^\circ}$.} First of all,
the terminology of `Reality' in the `Reality Problem of Measurement Outcomes' is inspired by the fact that 
if we describe the measurement interaction unitarily,
as in the proof above, none of the \emph{possible} 
measurement outcomes becomes \emph{real} --- 
in general, possession of a determinate
property by $\sfS$ and calling that property of $\sfS$ \emph{real}, or calling it \emph{actual}, is saying exactly 
the same with different words.\fn{To call
only possessed properties {$\la A,a\ra$} `determinate' leaves one without terminology for such properties when they are
not possessed. Not a good thing. We call them: mere 
possible but not actual determinate properties.} Then 
non-actual properties can be determinate, but are not real. 

\tbf{2$\mathbf{^\circ}$.} Notice that \emph{probability} is  not even mentioned 
in the Proof. Therefore, no matter how one
interprets probability in QM (`quantum probabilities'), 
this will never solve the Reality Problem of Measurement Outcomes. Also replacing commutative Kolmogorovian Probability Theory with non-commutative `Quantum Probability'
Theory is of no avail when it comes to the Reality Problem
of Measurement Outcomes.

\tbf{3$\mathbf{^\circ}$.} 
Requirement \eqref{Um} on measurement interactions is an instance of the
\display{$\blacksquare$~~\tbf{Property Revealing Condition.} \emph{If the pure state of the composite sysem 
$\,\sfS\sqcup\sfM\,$
is \mbox{$\,|a\ra\otimes|m_{0}\ra\in\calH_{\sfS}\otimes\calH_{\sfM}$}, where $|a\ra$ is an eigenvector of measured magnitude
$A$ of physical system $\sfS$, so that $\sfS$ has 
determinate property $\la A,a\ra$,
and if the measurement device $\sfM$ measures $A$ 
by operator $M$, and the unitary measurment interaction
is $\,t\mapsto U_{\mm}(t)$, then 
after the measurement has ended, at time $\,t=\tau$,
the state of $\,\sfS\sqcup\sfM\,$ is such that $\sfM$
reveals that property whilst $\sfS$ may have lost it
(the state is then an eigenvector of $\,\one\otimes M\,$);
below $\,|\phi\ra\in\calH_{\sfS}\,$ is any state of $\sfS$ and
$\,|m_{a}\ra\in\calH_{\sfM}\,$ is the state of $\sfM$ that
correlates with $\,|a\ra\in\calH_{S}$}:}
\protect\vspace*{-3ex}
\beq
U_{\mm}(\tau)\big(|a\ra\otimes|m_{0}\ra\big)\,=\,
|\phi\ra\otimes|m_{a}\ra\,.
\label{PropRevCond2}
\enq
When $\,|\phi\ra=|a\ra$, one speaks of an
\emph{ideal measurement}: the state $|a\ra$ of 
$\sfS$ is left 
undisturbed and $\sfS$ still has the determinate property
$\la A,a\ra$ in the post-measurement state \eqref{PropRevCond2} that 
it had in the inital state.
When $\,|\phi\ra\neq|a\ra$, one speaks of
a \emph{disturbance measurement}:
in the post-measurement state, $\sfS$ will then 
have lost property $\la A,a\ra$,
due to the measurement interaction. (You read your weight while standing on scales, leave the scales, and then have lost your weight --- Quantum Weight Watching.) In the proof
above, we assumed that the measurement interaction
was ideal, leading to requirement~\eqref{Um}.
The proof remains intact when we consider
disturbance measurements, as we shall point out
next.

Applied to the Stern-Gerlach experiment, 
we then have for the final states of $\sfe\sqcup\sfM$:
\beq 
U_{\mm}(\tau)\big(\zop\otimes|m_{0}\ra\big)\,=\,|u\ra\otimes|+\ra
\quad\trm{and}\quad
U_{\mm}(\tau)\big(\zneer\otimes|m_{0}\ra\big)\,=\,|v\ra\otimes|-\ra\,,
\label{Umd}
\enq   
where $\,|u\ra,|v\ra\in\C^{2}\,$ can be any states of $\sfe$.
Then the post-measurement state of $\,\sfM\sqcup\sfe\,$
is not \eqref{psitau} but becomes:
\beq \alpha|u\ra\otimes|+\ra\;+\;\beta|v\ra\otimes|-\ra\,,
\label{psitau2}\enq
which is neither an eigenstate of $\,\sigma_{z}\otimes M\,$,
nor of $\,\one\otimes M$,
and therefore $\sfM$ does not indicate an outcome.
The logical clash with SMOP remains within deductive reach
when we replace ideal with disturbance measurements.

Since $\,|v\ra,|u\ra\in\C^{2}$, they are superpositions
of $\zop$ and $\zneer$ (standard basis of $\C^{2}$). 
This implies that the 
disturbed post-measurement state \eqref{psitau2}
has terms $\,\zop\otimes|-\ra\,$ and 
$\,\zneer\otimes|+\ra\,$, suggesting that the
coefficients in front of these terms yield the probability
of $\sfM$ indicating the \emph{wrong} outcome (`false positives' and `true negatives').

\tbf{4$\mathbf{^\circ}$.} 
If standard QM were to include all premises mentioned in the 
Reality Problem of Measurement Outcomes, then standard QM would be inconsistent. Standard QM escapes the inconsistency argument narrowly because it rejects the Universal Dynamics Postulate (which implies that measurement interactions are unitary), and replaces it with a conditional version:
\display{\tbf{\bfbox~~Dynamics Postulate.}
\emph{Time is represented by the real continuum $(\R)$.
IF no measurements are performed on physical system $\sfS$
during time interval $\,I\subseteq\R$,
~THEN there is some connected continuous Lie-group
of unitary operators $\,t\mapsto U(t)\,$ acting on $\calH$
such that, when $|\psi(0)\ra$
represents the state at time $t=0$, the state at every time $\,t\in I\,$ is:}}
\vspace*{-3ex} 
\beq\,U(t)|\psi(0)\ra=|\psi(t)\ra\,.\label{Ut}\enq
This raises, of course, the question what happens 
when a measurement is 
performed. For that case, Von~Neumann advanced
another conditional 
postulate, such that the two postulates are mutually exclusive and jointly 
exhaustive:
\display{\tbf{\bfbox~~Projection Postulate.} \emph{IF
one performs a measurement of physical magnitude represented by operator $A$ on physical system $\sfS$, 
when $\sfS$ has state $\,|\psi(t)\ra\in\calH\,$ at the moment $t\in\R$ of measurement,
AND one finds spectrum value in interval $\,\Delta\subset\R\,$ as the outcome ($\Delta$ being the measurement accuracy), ~THEN immediately after this measurement outcome 
has been obtained, the post-measurement state of the physical system is $\,P^{A}(\Delta)|\psi(t)\ra$, where $P^{A}(\Delta)$
is the projector that projects onto the eigen-subspace
$\,\calH(A,\Delta)$.}}
The phrase `immediately after' can be made 
mathematically precise in terms
of upper and lower limits, but we gloss over this.

Most interpretations of QM reject a different premise of the ones mentioned in the Reality Problem of Measurement Outcomes to avoid inconsistency. The Copenhagen Interpretation
follows standard QM by adopting the Projection Postulate
and amending the Universal Dynamics Postulate.
Everett and Many Worlds reject
SMOP. Rovelli's Relational QM \emph{somehow} amends the Eigenlink. Modal Interpretations reject (one conjunct of)
the Eigenlink. Bohmian Mechanics adopts a stronger
state postulate, an additional postulate for worldlines of 
particles, and an involved story about measurements (reducing 
them all to position-measurements); it escapes the contradiction by never having superpositions of worldlines. Spontaneous 
collapse interpretations prevent repugnant superpositions of 
states of macroscopic physical systems to occur
by replacing the Dynamics Postulate with a different one,
positing some non-linear, and hence non-unitary
change of state over time.
\section{{\punt}Comparison to Maudlin's Three Measurement Problems}
\label{SectComp}
We analyse Maudlin's well-known three measurement problems.

\tbf{Problem~1.} 
Maudlin (1995) discerned three measurement problems and
one of them closely resembles the polylemma we have callded `The Reality Problem of Measurement Outcomes'
(Maudlin: ``Problem~1: the problem of outcomes'').
Maudlin took the State, Magnitude and Spectrum Postulate
for granted and did not even care to mention them;
he showed the inconsistency between
the following three ``claims'' (our italics):
\display{\small 1.A~~The wave-function of a system is \emph{complete}, i.e.\ the wave-function specifies (directly or indirectly) all of the physical properties of a system.\exmeer
1.B~~The wave-function always evolves in accord with
a \emph{linear} dynamical equation (e.g.\ the Schr\"{o}dinger equation).\exmeer
1.C~~Measurements of, e.g., the spin of an electron always (or at least usually) have \emph{determinate} outcomes, i.e., at the end of the measurement the measuring device is either in a state which indicates spin up (and not down) or spin down (and not up).}

Claim~1.A asserts that the state of every physical system $\sfS$ must \emph{somehow} yield \emph{all} determinate properties of $\sfS$. We point out that the Eigenlink states a criterion that precisely achieves this. Hence Claim~1.A is more general: the Eigenlink implies Claim~1.A, but Claim~1.A does not imply the Eigenlink. The \emph{completeness} of the state must be taken to imply that only the state and nothing else, autonomously determines what the possessed determinate properties are, notably \emph{not} in combination with the measurement context.

Claim 1.B follows from the Universal Dynamics Postulate, because unitary operators are linear mappings on Hilbert-space. Yet 1.B says more generally that the function $\,t\mapsto|\psi(t)\ra\,$ governing the change of state over time is `linear'. 

Claim 1.C equates (i) $\sfM$ showing a determinate outcome to (ii) $\sfM$ being in a relevant eigenstate. This claim is a terse combination of SMOP, the Spectrum Postulate and the Eigenlink, which three distinct propositions ought to have to 
been unsnarled. Maudlin writes (1995, p.$\,$8):
\display{\small So if 1.A and 1.B are correct, 1.C must be wrong. If 1.A and 1.B are correct, $z$-spin measurements carried out on electrons in $x$-spin eigenstates will simply fail to have determinate outcomes.}
The post-measurement state $\,|\Psi(\tau)\ra$~\eqref{psitau} is a superposition of $\,\zop\otimes|+\ra\,$
and $\,\beta\zneer\otimes|-\ra\,$. To deduce that neither the measured system has spin-$z$ properties nor the measuring device displays the relevant outcomes in this state, one needs to assume that being in an eigenstate is necessary for the
possession of these properties, which is `half' of the Eigenlink. 

When starting to present his second measurement problem,
Maudlin informs the reader that he has taken the reader for a ride when expounding Problem~1:
\display{\small The three propositions in the problem of outcomes are 
not \emph{strictu sensu} [\emph{sic}] incompatible. We used a symmetry 
argument to show that $S^{*}$ [our $|\psi(\tau)\ra$~\eqref{psitau}] could 
not, if it is a complete physical description, represent a 
detector which is indicating `UP' but not `DOWN' or 
\emph{vice versa}. But symmetry arguments are not a matter of 
logic. Since we have not discussed any constraints on how the 
wave-function represents physical states, we could adopt a 
purely brute force solution: simply stipulate that the state 
$S^*$ represents a detector indicating, say, `UP'. Then 1.A, 
1.B and 1.C could all be simultaneously true.}
Maudlin buried his Problem$\,$1 right after having expounded it. A hidden premise in his argument was, peculiarly, some `symmetry assumption', having to do with the equal coefficients $1/\sqrt{2}$ in the spin-singlet state. Since we did not need such an assumption at all in our proof of the Reality Problem of Measurement Outcomes, Maudlin's proof cannot be the same as our proof. In the proof of inconsistency of Problem$\,2$ (another polylemma), this `symmetry assumption' is relaxed, and hopefully we shall have a proof of inconsistency \emph{stricto sensu}.

\tbf{Problem~2.} Maudlin (1995, p.$\,$11) carries on to present a resembling yet different treesome of claims,
which are also mutually inconsistent (our italics):
\display{\small 2.A~~The wave-function of a system is \emph{complete}, i.e.\ the wave-function specifies (directly or indirectly) all of the physical properties of a system.\exmeer
2.B~~The wave-function always evolves in accord with a \emph{deterministic} dynamical equation (e.g. the Schr\"{o}dinger equation).\exmeer
2.C~~Measurement situations which are described by \emph{identical} initial wave-functions sometimes have different outcomes, and the \emph{probability} of each possible outcome is given (at least approximately) by Born's rule.}

Claim~2.A is identical to Claim~1.A. The difference between Claim~2.B and claim~1.B is that the `linear' has been replaced
with `determinstic'. Both claims 1.B and 2.B follow from the Universal Dynamics Postulate but sting at different properties of the unitary evolution. We can therefore
restrict our attention to the only substantial 
difference between Maudlin's Problem~1 and Problem~2, which is Claim~2.C. 

The first conjunct of Claim~2.C asserts that the relation between wave-functions and measurement outcomes is 
\emph{not a function}, from $\calH$ to the spectrum of $A$,
for every physical magnitude $A$:
different measurement outcomes, different wave-functions.\fn{More precisely: not related by a global phase factor, so 
\emph{stricto sensu} a function from a partition of $\calH$
to the spectrum of $A$, with equivalence relation:
\mbox{$\,|\psi\ra\sim|\phi\ra\,$} iff there is some $\alpha\in\C$
such that \mbox{$\,|\phi\ra=\alpha|\psi\ra$}.} Let's call this a \emph{specification function}, in consonance with 
the terminology of Claim~2.A (below $[\calH]_{\sim}$ is the partition of rays, and Sp$(A)$ is the spectrum of~$A$):
\beq
f_{A}:[\calH]_{\sim}\to\;\trm{Sp}(A),\;
[\psi]\mapsto f_{A}(\psi)\,. \label{fA}\enq
Of course there exists an infinitude of specification functions in the mathematical domain of discourse, but Claim~2.A asserts that \emph{one} of these functions is somehow `realised in nature'; this function represents a relation in physical reality, just as each moment in time, \emph{one} Hilbert-vector (better: one ray) of the non-denumerably many Hibert-vectors represents the state of a physical system, and all others do not represent the state at that moment. Claim~2.A essentially calls \emph{this} the `completeness' of the wave-function.

The second conjunct of Claim~2.C involves the Probability Postulate of QM:
\display{\tbf{\bfbox~~Probability Postulate.}
\emph{Suppose we perform a measurement on a physical system
$\sfS$ of physical magnitude (represented by self-adjoint operator) $A$ at time {$\,t\in\R\,$} while the physical state 
of the system is represented by {$\,|\psi(t)\ra\in\calH$}. Then the probability of finding upon measurement some value in Borel set {$\Delta\in\calB(\R)$} is given by the Born probability measure:}}
\vspace*{-3ex}
\beq 
{\Pr}^{|\psi(t)\ra}(A:\Delta)\,=\,
\la\psi(t)|P^{A}(\Delta)|\psi(t)\ra\,.
\label{PrA}\enq

Problem~2 is indeed different from Problem~1, and 
different too from our Reality Problem of Measurement Outcomes, precisely because it involves probabilities.
Maudlin (\emph{ibid.}):
\display{\small The inconsistency of 2.A, 2.B and 2.C is patent: If 
the wave-function always evolves deterministically (2.B), 
then two systems which begin with identical wave-functions 
will end with identical wave-functions. But if the 
wave-function is complete (2.A), then systems with identical 
wave-functions are identical in all respects. In particular, 
they cannot contain detectors which are indicating different 
outcomes, contra 2.C.}

If the state (wave-function) determines the determinate properties \emph{probabilistically}, then the same state
does not specify \emph{which} determinate properties are possessed, and
Maudlin's argument collapses. The argument is valid
if, and only if, the \emph{completeness} of the state (2.A) 
is supposed to entail that the state specifies all possessed determinate properties \emph{non-probalistically},
or \emph{determines} them, say, so that the same state 
always specifies the same possessed determinate properties, as Maudlin says (1995. p.$\,$11), and as e.g.\ the Eigenlink ordains. The Eigenlink provides a specification function \eqref{fA} for every $A$:
\beq
f_{A}(\psi)|\psi\ra\,=\,A|\psi\ra\,.
\label{fAEigen}\enq

The inconsistency of Maudlin's Problem~2 occurs already
between the Probability Postulate (second conjunct of Claim~2.C), and the weaker claim 
that every state {$\,|\psi\ra\in\calH\,$} of system $\sfS$ is
\emph{measurement-complete}, which is to say that 
$|\psi\ra$, in combination with an ideal 
measurement apparatus ($\sfM$) in some initial state, 
determines a unique measurement outcome $m_{a}$ for every 
physical magnitude $A$, correlated to value $a$ from the spectrum of $A$ (call it claim 2.A$'$). 
A relevant specification function $g_{A}$ can be introduced
that sends, for each $A$ and $A$-measuring operator $M$ of $\sfM$, rays in {$\calH_{\sfS}\otimes\calH_{\sfM}$}
to ordered pairs of values from the spectra of $A$ and $M$:
\beq
g_{A}:[\calH_{\sfS}\otimes\calH_{\sfM}]_{\sim}\to\;\trm{Sp}(A)\times\trm{Sp}(M),\;[\Psi]_{\sim}\mapsto g_{A}(\Psi)=\la a,m_{a}\ra\,.
\label{gA}
\enq
Or perhaps only a specification function for $M$,
similar but not identical to $f_{A}$~\eqref{fA}:
\beq
m_{A}:[\calH_{\sfS}\otimes\calH_{\sfM}]_{\sim}\to\;\trm{Sp}(M),
\;[\Psi]_{\sim}\mapsto m_{A}(\Psi)=m_{a}\,.
\label{mA}
\enq

Consider a state that is a 
superposition in the measurement basis and we
are done: the measurement outcome should always be the same due to assumed the measurement-completeness of
the state, which is contradicted by the 
Probability Postulate because \emph{it} 
generically gives non-zero
probabilities for other measurement outcomes
whilst the system is in the same state. The addition of the
Universal Dynamics Postulate (which implies Claim~2.B) is logically superfluous. Claim~2.A, expressing the 
\emph{property-completeness} of every state, 
implies 2.A$'$, which is only about measurement outcomes and
not about possessed determinate properties.
We can strenghten Maudlin's Problem~2 as the
\display{\donkrood{\ding{115}}~~\tbf{II. 
State Completeness Problem.}
\emph{Given the State, Magnitude and Spectrum Postulate.
Then the Probability Postulate is incompatible with
the measurement-completeness as well as the 
property-completeness of the state.}
\label{StateCompl}}

The absence of the deterministic character of the change of state over time, as in the Universal Dynamics Postulate, and of any postulate about how states change over time for that matter, makes the State Completeness Problem irrelevant for the question whether or not measurement interactions are unitary or not. 

But the situation of Problem~2 is logically even worse --- or better? Claim~2.A, stating the property-completeness of the  state, by asserting the realisation in nature of \emph{some} specification function $f_{A}$~\eqref{fA}, \emph{almost} contradicts the first conjunct of Claims~2.C, which denies that the relation between states and measurement outcomes is a function; this comes down to denying the realisation in nature of \emph{any} specification function $g_{A}$~\eqref{gA} or $m_{A}$~\eqref{mA}. \emph{Almost} contradicts, we wrote, because to obtain a contradiction, we only have to add that the specification functions are consistent: 
\beq
\textrm{if ~{$|\Psi\ra\,=\,|\psi\ra\otimes|\phi\ra\,$},
~then~ $g_{A}(\Psi)=\la f_{A}(\psi),\;m_{A}(\phi)\ra\,$.}
\label{fAM}
\enq
Besides the Dynamics Postulate (Claim~2.B), even the Probability Postulate drops out now (second conjunct of Claim~2.C). Otiose. We shall not elevate this inconsistency to another polylemma  `measurement problem', forcing one to choose between the State, Magnitude and Spectrum Postulate, and the consistency of the specification functions \eqref{fAM}, because it is silly  to state the property-completeness of the state in one premise (Claim~2.A) and nearly to deny it in another premise (first conjunct of Claim~2.C).

Since standard QM includes all of the four mentioned postulates, the State Completeness Problem implies that QM is measurement- as well as property-incomplete, which points into the direction of indeterminism. This is similar to but \emph{not the same as} the incompleteness conclusion of Einstein, Podolsky and Rosen (1935), but now reached without having to assume any locality condition or to employ entangled states of two particles, and, trotting in the footsteps of Fine (1986, Ch.$\,$3), perhaps even closer to Einstein's intentions.\fn{Traditionally,  `the completeness problem' is whether there is another  theory, a `hidden-variables theory', with additional degrees of freedom, that performs empirically just as good as standard QM but is not haunted by the problems presented in this  paper. See Bub (1974, Ch.$\,$II).} \emph{Not the same as}, we wrote, because for EPR, completeness can only be established by first knowing which determinate properties are possessed by means of their reality condition (`elements of physical reality'), and then inquiring into whether QM permits or forbids their possession.

The measurement- and state-incompleteness is only a \emph{problem} for determinists.  Friends of standard QM, ready accept indeterminism governing physical reality at the scales of the tiny and the brief, will see nothing problematic about the  ~\donkrood{\ding{115}}~State Completeness Problem; they will take it as an expression of the indeterministic character of QM.

\tbf{Problem~3.} Maudlin's third measurement problem 
(``the problem 
of effect'') concerns some subspecies of one species of
interpretation of QM, namely the Modal Interpretation,
and it concerns repeated measurements. In a nutshell, the problem is 
that the measurement outcome of one measurement, which 
reveals a possessed determinate property of the measured system 
according 
to Modal Interpreters, \emph{has no effect} on subsequent 
measurement results, which also reveal possessed 
determinate properties. Maudlin 
does not deduce a contradiction from explicit claims, and 
therefore \textbf{Problem~3} is not in the same logical category as his other two problems.\fn{G.\ Bacciagaluppi 
has suggested that the modal interpretation is a red herring in Problem~3, and that Problem~3 points at the general
issue of repeated measurements, which could be elevated to a seventh measurement problem. Private communication, Utrecht, October 2022.}

To recapitulate, Maudlin burried \tbf{Problem~1}, but with some tweaking, and dispensing with his peculair and surreptitious `symmetry assumption', it becomes the Reality Problem of Measurement Outcomes (p.~\pageref{RPMO}). We could improve on Maudlin's \tbf{Problem~2} by arriving at a contradiction between fewer premisses; in fact between, granted a few uncontroversial postulates of standard QM: the Probability Postulate and the claim that the state is measurement-complete, which is implied by being property-complete (State Completeness Problem). Both \tbf{Problem~2} and \tbf{Problem~1} are problems that compel one to reject at least one of a number of premises. \tbf{Problem~1} is not a problem \emph{of standard} QM due to its Projection Postulate, and \tbf{Problem~2} (State Completeness Problem) can be seen as a simple \emph{proof} in QM of the measurement- and property-incompleteness of the state, expressing its indeterministic character --- only a \emph{problem} for determinists. \tbf{Problem~3} is not a problem for standard QM, but for a subspecies of the Modal Interpretation of QM: ultimately it states the open problem of finding a dynamics of possessed properties, in light of the fact that measurement-outcomes seem to be irrelevant for subsequent property ascriptions in most modal interpretations.

We move on to four other measurement problems.
\section{{\punt}The Probability Problem of Measurement Outcomes}
\label{SectInsolu}
\tbf{Introduction.}
So-called `Insolubility Theorems' suggest 
that the Reality Problem of Measurement Outcomes is
insoluble. Wrong suggestion. These theorems are only about probability distributions of measurement outcomes
when the measurement interaction is taken to be unitary,
as implied by the Universal Dynamics Postulate.
What the Reality Problem of Measurement Outcomes has 
in common with the Probability Problem of Measurement
Outcomes (\emph{lege infra}) is that both make trouble for taking measurement
interactions to be unitary. A difference is that
this new Probability Problem crucially involves mixed states
and probability --- both are absent from
the Reality Problem of Measurement Outcomes.
The core of the insolubility proof goes back straight to 
Von~Neumann's discussion of the measuring proces, in 
Section~VI.3 of his \emph{Grundlagen} (1932). We first expound this 
core as applied to the same Stern-Gerlach experiment we used 
in the proof of the Reality Problem of Measurement Outcomes.
Then we ascend to levels of utmost generality. 

\tbf{Core and Special Case.}
The State Postulate needs to be extended from Hilbert-vectors
representing pure physical states to \emph{state operators},
aka \emph{density operators},
which are by definition 
self-adjoint, positive, trace~$1$ operators,
collected in convex set $\calS(\calH)$. On the boundary
of this set, one finds $1$-dimensional projectors, which are the pure states because they correspond one-one to 
(rays of) Hilbert-vectors:
$\,|\phi\ra$ and $\,P_{\phi}=|\phi\ra\la\phi|$. 
The Probability Postulate generalises from \eqref{PrA}
to Von~Neumann's celebrated trace-formula, for
state operator $W\in\calS(\calH)$:
\beq 
{\Pr}^{W}(A:\Delta)\,=\,\trm{Tr}\big(WP^{A}(\Delta)\big)\,.\label{TrWP}\enq
The Projection Postulate also generalises to mixed states,
given by L\"{u}ders' formula; but we shall not need it here
and therefore gloss over it.\fn{Bush \emph{et al.} (1996, 
pp.$\,$31, 40--41).}

We consider the Stern-Gerlach experiment again.
Suppose the initial state of the electron ($\sfe$) is a pure
state: 
\beq W_{\sfe}(0)\,=\,P^{z}_{\uparrow}\,\equiv\,\zop\langle\uparrow\!|\,.
\label{We0}\enq
Suppose the initial 
state of the measurement device ($\sfM$) is mixed; we write
it as a convex combination of orthogonal pure states, each of 
which projects on an eigenvector 
of the measuring operator $M$:
\beq W_{\sfM}(0)\,=\,w_{0}P^{M}_{0}+w_{1}P^{M}_{+}+
w_{2}P^{M}_{-}\,.
\label{WM0}
\enq
where the real coefficients $\,w_{j}\in [0,1]\,$ sum up to~$1$. Combination \eqref{WM0} is unique by the Spectral Theorem.

The ideal measurement interaction $U_{\mm}(t)$ then leads to 
the following final state (at time $\,t=\tau$), using
\eqref{We0} and \eqref{WM0}:
\beq
\begin{array}{lcl}
W(\tau)&=& U_{\mm}(\tau)W(0)U_{\mm}^{\dagger}(\tau)\exmeer
&=& U_{\mm}(\tau)\big(W_{\sfe}(0)\otimes W_{\sfM}(0)\big)U_{\mm}^{\dagger}(\tau)\exmeer
&=&w_{0}\,U_{\mm}(\tau)\big(P^{z}_{\uparrow}\otimes P^{M}_{0}\big)U_{\mm}^{\dagger}(\tau)\;+\;w_{1}\,
U_{\mm}(\tau)\big(P^{z}_{\uparrow}\otimes P^{M}_{+}\big)U_{\mm}^{\dagger}(\tau) \exmeer
& &\;+\;w_{2}\,
U_{\mm}(\tau)\big(P^{z}_{\uparrow}\otimes P^{M}_{-}\big)U_{\mm}^{\dagger}(\tau)
\end{array}
\label{WeMtau}
\enq
Being a projector is invariant under unitary transformations. Since \mbox{$\,P^{z}_{\uparrow}\otimes P^{M}_{k}\,$} are projectors on \mbox{$\,\calH_{\sfS}\otimes\calH_{\sfM}$}, the final state is a convex combination of orthogonal pure states corresponding to vectors (cf.\ footnote~\pageref{fnmj}, p.$\,$\pageref{fnmj}):
\beq U_{\mm}(\tau)\big(\zop\otimes|m_{j}\ra\big)\,.
\enq

Enter \emph{ensembles}.\fn{Neumann (1927, 1932).} Suppose we have large number of copies of composite systems \mbox{$\,\sfe\sqcup\sfM$}, and we want to characterise
ensembles by mixed state operators. Suppose further that
every copy of $\sfe$ initially is the same pure $z$-spin-state, characterised by $P^{z}_{\downarrow}$ or by $P^{z}_{\uparrow}$. Such an ensemble is called \emph{homogeneous}, and is characterised by some pure state operator such as $W_{\sfe}(0)$~\eqref{We0}. Every copy of $\sfM$ in the ensemble is also in some pure state, but we assume not in the same one, and we do not know in which one. 
Such an ensemble is called \emph{heterogeneous},
and characterised by $W_{\sfM}(0)$~\eqref{WM0}.
The coefficient $w_{j}$ is the probability
that a copy of $\sfM$ is in pure state $P^{M}_{j}$,
in agreement with the trace-formula \eqref{TrWP}:
\beq
\trm{Tr}\left(W_{\sfM}(0)P^{M}_{j}\right)\,=\,
\sum_{k=0}^{2}w_{k}\delta_{kj}\,=\,w_{j}\,.
\label{TrWM0PM}
\enq 

This is called the \emph{ignorance interpretation of mixtures}. \label{IgnMixtures}
Can we now also interpret final state $W(\tau)$ \eqref{WeMtau}
in this fashion? That is, every copy of 
\mbox{$\,\sfe\sqcup\sfM\,$}
is in a pure state \mbox{$\,P^{z}_{\uparrow}\otimes P^{M}_{j}$}, and hence by the Eigenlink, $\sfe$ possesses determinate property \mbox{$\la\sigma_{z},\uparrow\ra$}, and $\sfM$ possesses accompanying determinate property {$\la M,m_{k}\ra\,$}? Our ignorance about the initial state of
a copy of $\sfM$ in the ensemble is preserved as the same 
ignorance about the final state of that copy of~$\sfM$.

Let us now consider a different initial pure state of $\sfe$:
\beq 
|\phi(0)\ra\,=\,\alpha\zop+\beta\zneer\;\in\calH_{\sfe}=\C^{2}\,. 
\label{phi0}
\enq 
The initial state of the ensuing ensemble of copies of the 
composite system $\,\sfe\sqcup\sfM\,$, where we retain the
same mixed state $W_{\sfM}(0)$ for $\sfM$ as 
before \eqref{WM0}, then is:
\beq 
W(0)\,=\,|\phi(0)\ra\la\phi(0)|\otimes W_{\sfM}(0)\,.
\label{WeM0}
\enq
The final, post-measurment state becomes:
\beq
W(\tau)\,=\,\sum_{j=0}^{2}w_{j}\,
U_{\mm}(\tau)\left(|\phi(0)\ra\la\phi(0)|\otimes P^{M}_{j}\right)U_{\mm}^{\dagger}(\tau)\,,
\label{WeMtau2}
\enq
which is a heterogeneous mixture of three pure states:
\beq
U_{\mm}(\tau)\left(P_{|\phi(0)\ra}\otimes P^{M}_{j}\right)U_{\mm}^{\dagger}(\tau)\,.
\enq

Remarkably, the probability that measurement outcome $m_{j}$ obtains equals again $w_{j}$, just as in the initial mixed state $W_{\sfM}(0)$~\eqref{WM0}, as expressed in \eqref{TrWM0PM}. This probability does not depend on the initial state of $\sfe$: the coefficients $\,\alpha,\beta\in\C$, charactarising the initial pure state of $\sfe$ \eqref{phi0}, are \emph{absent} from \eqref{WeMtau}, as also 
expressed in \eqref{TrWM0PM}. This is in conflict
with a second requirement on $U_{\mm}(t)$ to qualify as a 
measurement interaction, a condition that involves probabilities, which we did not need in the Reality Problem of Measurement Outcomes. We shall state it below in full 
generality.\fn{Bush \emph{et al.} (1996, p.$\,$29).}

\tbf{General Case.}
First some general stage setting.  
We consider a physical system $\sfS$, 
a measurement device $\sfM$, their composite system
\mbox{$\,\sfS\sqcup\sfM$}, and their sets of mixed states
$\,\calS(\calH_{\sfM})$, $\,\calS(\calH_{\sfS})\,$ and 
\mbox{$\,\calS(\calH_{\sfS}\otimes\calH_{\sfM})$}, respectively. Physical magnitude $A$ of $\sfS$ we take to be self-adjoint. We subdivide the scale $\Delta_{M}$ 
of $\sfM$, which is the spectrum
of $A$-measurement operator $M$, in $N$ intervals {$\,I_{j}\subset\Delta_{M}$}, and $m_{j}$ being the 
midpoint of $I_{j}$; the equal with of $I_{j}$
coincides with the measurement accuracy. When we include the ready-to-measure pure state of $\sfM$, then $N+1$ orthogonal states of $\sfM$ suffice, which means that $\calH_{\sfM}$
is finite-dimensional, with dimension $N+1$. 
Suppose we can measure part $\Delta_{A}$ of the
spectrum of $A$, perhaps even of the entire spectrum
of~$A$. A \emph{calibration function} 
sends spectrum-values of $A$ to measurement outcomes (spectrum-values of $M$):
\beq
g:\Delta_{A}\to\Delta_{M},\;a\mapsto g(a)
\enq
One assumes $g$ to be one-one and continuous, so that $g$ correlates values in $\Delta_{A}$ to values in $\Delta_{M}$ perfectly. Just as the $N$ intervals $I_{j}$ partition measurement scale $\Delta_{M}$, intervals 
\mbox{$\,g^{\mrm{inv}}(I_{j})\subset\Delta_{A}\,$} partition 
part $\Delta_{A}$ of the spectrum of~$A$. The unitary measurement evolution $\,t\mapsto U_{\mm}(t)\,$ sends the initial mixed state $W(0)$ to the mixed final state:
\beq
W(\tau)\,=\,U_{\mm}(\tau)\big(W_{\sfS}(0)\otimes W_{\sfM}(0)\big)U_{\mm}^{\dagger}(\tau)\,.
\label{WSMtau}
\enq 
Hence we arrive at the:
\display{$\blacksquare$~~\tbf{Probability Reproducibility Condition.}
\emph{The Born-Von~Neumann probability measure \eqref{TrWP} 
for $A$ in the initial state of $\sfS$ is the same 
as the probability measure of $\sfM$ for $M$ in the final 
state of the composite system $\,\sfS\sqcup\sfM\,$ when
unitarily evolved by} $U_{\mm}$~\eqref{WSMtau}:}
\protect\vspace*{-3ex}
\beq
{\Pr}^{W_{\sfS}(0)}(A:\Delta_{j})\,=\,
{\Pr}^{W(\tau)}\big(\one\otimes M:g^{\mrm{inv}}(\Delta_{j})\big)\,.
\enq

For the Stern-Gerlach case, we then must have, for 
arbitrary pure initial state $|\phi(0)\ra$ 
of $\sfe$ \eqref{phi0}, using \eqref{WeM0}
and \eqref{WeMtau2}:
\beq 
w_{1}=|\alpha|^{2}\,,
\quad w_{2}=|\beta|^{2}\quad
\trm{and}\quad w_{0}=0\,.\label{w12}
\enq
So for every initial mixed state $W_{\sfM}(0)$ \eqref{WM0}
with coefficients $w_{1}$ and $w_{2}$ different from
$|\alpha|^{2}$ and $|\beta|^{2}$, respectively, we
have a logical clash with the Probability Reproducibility
Condition via eqs.~\eqref{w12}.

This is essentially a proof of the core of the:\fn{Cf.\ Theorem~6.2.1 in
Bush \emph{et al.} (1996, p.$\,$76).}
\display{\donkrood{\ding{115}}~\tbf{III. Probability Problem of Measurement Outcomes.} \emph{Granted
the Mixed State Postulate and the Magnitude Postulate. 
Then the Probability Postulate, the 
Universal Dynamics Postulate, and the
Probability Reproducibility Condition are jointly
incompatible.}\label{PPMO}}

A few supplementary remarks about this polylemma problem.

\tbf{(a)}~~Notice that \emph{not} among the six jointly inconsistent premises
are: the Spectrum Postulate, the Eigenlink and SMOP,
which are members of the inconsistent bouquet of
the Reality Problem of Measurement Outcomes.

\tbf{(b)}~~What Von~Neumann pointed out (\emph{lege supra}) is the
core of the proof. Wigner critically discussed Von~Neumann's considerations and repeated the core (1963, p.$\,$12). Fine (1970) fashioned it into a strengthened theorem with a proof.
Fine's proof sadly was ``seriously defective'', as Stein (1997, p.$\,$233) would put it; Shimony (1974) 
performed a repair job.
Bush and Shimony (1997) extended the Insolubility Theorem from
self-adjoint operators to positive operators (`unsharp'
physical magnitudes). 
Brown (1986, p.$\,$862) claimed to provide ``a simple and transparent proof'', but entangled it with the ignorance interpretation of mixtures and privileged convex expansions.
Stein (1997, p.$\,240$) flogged Brown for this:
\display{\small This simple proof in question is not a proof 
of the theorem I have presented here, or of the theorem
demonstrated by Shimony; nor is it a proof of the theorem
stated by Fine. What it establishes is something very much
weaker --- which, however, Brown maintains, is the only thing
that genuinely bears on the problem of measurement.}
Stein (1997, pp.$\;$240--241) ends as follows:
\display{\small In other words, Brown's proposal is the one 
already discussed in Section~1, above. Setting aside any 
questions about the viability of the notion of the ``real 
mixture'' --- the notion, that is, that a quantum statistical 
state should be characterized by more than its assignment of 
probabilities to values of dynamical variables and, in 
particular, that such a state should be thought of as an 
assignment of something like probabilities to pure states --- it 
has there been pointed out that with such a conception of the 
state it is trivial that appeal to the mixed initial state of 
the apparatus can contribute nothing to the measurement problem. 
It would hardly have been necessary for such a man as Wigner to 
undertake an examination of the question.}

\tbf{(c)}~~The proposal Brown rules out with his `Insolubility 
Theorem', 
and what Stein discusses and dismisses in his introductury 
Section~1, is the impossibility of a unitary measurement 
interaction 
such that the final, post-measurement state is a mixture
of pure states of the measuring magnitude $M$ to which
an `ignorance interpretation of mixtures' can be applied.

Recall that the ignorance interpretation of mixtures is the idea that when we write $\,W\in\calS(\calH)\,$ as some convex
combination of pure states:
\beq W\,=\,\sum_{n=0}^{N}w_{n}P_{n}\,, \enq
we should think of $W$ characterising an ensemble,
each member of which is in a pure state $P_{n}$ with
probability $w_{n}$. When we choose for $P_{n}$
orthogonal members of the spectral resolution of magnitude $A$
having a discrete spectrum,
then the Eigenlink permits us to say that each member of the
ensemble has determinate property $\,\la A,a_{n}\ra$, where
$P_{n}$ then projects on eigensubspace $\calH(A,a_{n})$.
Call such a convex expansion an $A$-\emph{expansion}.

We can also choose a $B$-expansion for $W$ such that
$B$ does not commute with $A$. Since non-commuting operators 
generically have no common eigenvectors, an ignorance 
interpretation of $W$ in terms of pure eigenstates of 
both $A$ and of $B$ is impossible.

To save the ignorance interpetation of mixtures, 
one can \emph{privilege} certain convex expansions and permit
an ignorance interpretation of only those privileged ones.
This is essentially what Brown (1986) does. Then Brown
further requires that \emph{if} the expansion of the 
initial state of the composite system $\,\sfS\sqcup\sfM\,$
is $M$-privileged:
\beq W(0)\,=\,\sum_{k=0}^{N}\,w_{k}\,P^{A}_{j}\otimes 
P^{M}_{k}\,,
\enq
where every copy of $\sfS$ is taken to be in pure 
state $P^{A}_{j}$, and $A$ is the
magnitude of $\sfS$ that $\sfM$ is measuring,
\emph{then} the final state has the following privileged
expansion:
\beq W(\tau)\,=\,\sum_{k=0}^{N}\,w_{k}\,U_{\mm}(\tau)\left(P^{A}_{j}\otimes P^{M}_{k}\right)U^{\dagger}_{\mm}(\tau)\,.
\enq
By considering empirically \emph{distinguishable} 
initial pure states of $\sfS$,
and showing that a final state ensues with weights 
identical to weights of the initial mixed state of $\sfM$, these
final states are empirically \emph{indistinguishable}, every
term being an eigenstate of $\,\one\otimes M$. 
This contradicts the Probability Reproducibility
Condition, but also the weaker condition that
empirically distinguishalbe initial states of $\sfM$
ought to lead by $U_{\mm}$ to empirically distinguishalbe
final states of $\sfM$.

Does the reliance of Brown's proof on $A$-privileged 
convex espansions as well as on an ignorance interpretation
of mixed states makes it diverge from the line of 
Insolubility Theorems that started with Fine
and originated in Von~Neumann (1932)?
We answer this question in the course of the next remark.

\tbf{(d)}~~Somewhat remarkable is that all proofs of Insolubility
Theorems start with a \emph{pure} state for $\sfS$ and a \emph{mixed}
state for $\sfM$. Brown (1986, p.$\,$863) indeed wonders
why one does not assume that $\sfM$ is prepared in a pure
intial state, the ready-to-measure state $|m_{0}\ra$.
Has the experimentator been drinking? Techo-House Party in the Laboratory with XTC? 

Is the pure ready-to-measure state $|m_{0}\ra$
the only proper initial state for $\sfM$? No.
Eigenvalue $m_{0}$ can be zillion-fold degenerate,
with zillion ($Z$) different joint states of the 
octillions of atoms 
that compose $\sfM$.
But then initial state $W_{\sfM}(0)$ can also be a mixture
of precisely these pure states, say $P^{0}_{n}$,
$\,n\in\{1,2,\ldots, Z\}$. There will be fluke terms in the convex expansion of $W_{\sfM}(0)$ in the 
$M$-basis,
with epsilonic probability. One then has:
\beq
W_{\sfM}(0)\,=\,W^{0}_{\sfM}(0)\;+\;\sum_{j=1}^{N}w_{k}P^{M}_{j}\,=\,\sum_{n=1}^{Z}v_{n}P^{0}_{n}\;+\;\sum_{j=1}^{N}w_{k}P^{M}_{j}\,, \enq
such that the sum of all
$w_{j}$ for $\,j\geqslant 1\,$ being equal to $\varepsilon$,
where $\,0<\varepsilon\ll 1$, and of all $v_{n}$
being $w_{0}$. Then, when $P^{M}_{0}$ projects
onto the eigenspace $\calH_{\sfM}(M,m_{0})$, we have:
\beq
P^{M}_{0}\,=\,\bigoplus_{n=1}^{Z}P^{0}_{n}\quad\trm{and}\quad
\trm{Tr}\big(W^{0}_{\sfM}(0)P^{M}_{0}\big)\,=\,1-\varepsilon\,\approx\,1\,.
\enq
Of course, starting with
this `realistic' initial state of $\sfM$ does not makes one
deviate from the collision course to the Probability Reproducibility Conditon. 

Contrastively, why restrict the initial state of
$\sfS$ to be pure? That seems an unnecessary restriction.

Suppose we were to start with a mixed
initial state of $\sfS$, and pure initial state $\,P^{M}_{0}\in\calS(\calH_{\sfM})\,$ of $\sfM$:
\beq W(0)\,=\,W_{\sfS}(0)\otimes W_{\sfM}(0)\,=\,
\sum_{k=0}^{N}\,p_{k}\,P^{A}_{k}\otimes 
P^{M}_{0}\,=\,\sum_{k=0}^{N}\,p_{k}\,P^{A}_{k}\otimes 
|m_{0}\ra\la m_{0}|\,.
\enq
Then the final state would be:
\beq W(\tau)\,=\,\sum_{k=0}^{N}\,p_{k}\,U_{\mm}(\tau)\big(P^{A}_{k}\otimes P^{M}_{0}\big)U^{\dagger}_{\mm}(\tau)\,.
\enq
Suppose we further were to impose condition \eqref{Um} of ideal measurements on $U_{\mm}(t)$, so that in terms of state operators:
\beq U_{\mm}(\tau)\big(P^{A}_{k}\otimes P^{M}_{0}\big)U^{\dagger}_{\mm}(\tau)\,=\,P^{A}_{k}\otimes P^{M}_{k}\,.
\enq
Then the final state makes one deviate from a collision course to the Probability Reproducibility Conditon:
\beq W(\tau)\,=\,
\sum_{k=0}^{N}\,p_{k}P^{A}_{k}\otimes P^{M}_{k}\,,
\enq
because this final state depends on the initial state
$W_{\sfS}(0)$ of $\sfS$. Then the probability for finding
$\sfM$ indicating $m_{k}$ is:
\beq {\Pr}^{W(\tau)}(M:m_{k})\,=\,
\trm{Tr}\big(1\otimes M)P^{M}_{k}\big)\,=\,p_{k}\,.
\enq
Safe at last! 

One can also start with both $\sfS$ and $\sfM$ in mixed
initial states (the most general case conceivable), and obtain, in case of ideal measurements,
a final state that has a convex expansion in pure
states $\,P^{A}_{j}\otimes P^{M}_{j}\,$ with
coeffients $p_{k}w_{j}$. Then for various choices of
these coefficients, one can collide again with the
Probability Reproducibility Condition. 

The conclusion is that problems only arise when the initial
state of $\sfM$ is mixed.

We are now in a good position to answer the question
posed at the end of the previous remark: 
Does the reliance of Brown's proof on $A$-privileged 
convex espansions as well as on an ignorance interpretation
of mixed states makes it diverge from the line of 
Insolubility Theorems that started with Fine
and originated in Von~Neumann (1932)?

As we have seen above in remark \tbf{(b}), Stein answered
harshly in the affirmative and trashed Brown's version. But this does not sit comfortably with Brown's motivation, which crucially involves 
the ignorance intepretation of mixtures (\emph{vide supra}, p.~\pageref{IgnMixtures}). This motivation is an attempt
to interpret the statistical spread in measurement outcomes of magnitude $A$, say, when every copy of $\sfS$ is prepared identically, as ignorance about the pure state each member of the ensemble of copies of $\sfS\sqcup\sfM$ is in after the
measurement, which in turn would reflect our ignorance about
the pure initial state of $\sfM$. This suggests that when
the initial state of $\sfM$ of the ensemble is mixed,
the initial state of the heterogeneous ensemble carries over the heterogenity of its final state, explaining the spread in measurement outcomes. Every copy of the ensemble would 
then be in a pure state acoording to the ignorance interpretation, and would have a determinate property $\la A,a_{j}\ra$. We would be on our way to dissolve the Reality Problem of Measurement Outcomes. Brown's (1986) is indeed titled `The Insolubility Proof of the Quantum Measurement Problem', and his requirement `RUE' (\emph{ibid.}, p.~860) then makes sense, which is that if the initial mixture is 
$A$-privileged (``the real mixture''), then the final, 
post-measurement mixture written in basis $\,U_{\mm}(\tau)|a_{j}\ra\la a_{j}|\,$ is the real one when the measurement interaction is unitary. \emph{Alas!} The Probability Problem of Measurement Outcomes now teaches us that this way of interpreting the statistical spread in identically prepared physical systems runs afoul against the Probability Reproducibility Condition. 
Apparently Stein had little patience for this
motivation: he considered the idea of an $A$-privileged
expansion (``the real mixture'' in Brown's words) 
as a non-starter.\fn{G.\ Bacciagaluppi insisted on making this point (prive communication, 7 October 2022).}

\tbf{(e)}~~Bacciagaluppi (2014) has recently pointed out that an insolubility theorem follows from the No-Signalling Theorem of QM. \emph{Wut?} The No-Signalling Theorem says that when two physical systems, $\sfS_{1}$ and $\sfS_{2}$ say, \emph{are not interacting}, the probability measure for $A$ of $\sfS_{1}$ (of outcomes) does not depend on \emph{which} magnitude one measures on $\sfS_{2}$ (settings), and \emph{vice versa}. This is also known as \emph{outcome-setting independence}. In case of modelling measurement interactions unitarily, $\sfS$ and  $\sfM$ however do interact, and one of them measures the other. So it seems we have two rather different situations on our hands, a non-interacting and an interacting composite system, giving rise to the same statement of independence. The reason for this independence then must be different in each case. And it is: the non-interaction makes the relevant joint probability measures factorise versus the conditions on the unitary interaction to qualify as a measurement interaction; cf.\ Bacciagaluppi (2014).

\tbf{(f)}~~Stein (1997) has claimed to 
provide ``the maximal extension'' of the Insolubility
Theorem: the weakest assumptions and the largest reach.
The issue Stein is concerned with is 
whether there always is a convex expansion of the final
state in terms of pure states that are eigenstates of the 
measuring magnitude~$M$ whenever the initial state is
thusly expanded.  
Since commuting operators share their eigenvectors (if they have any), the requirement on the unitary interaction
$U_{\mm}(t)$ to qualify as a measurement interaction just mentioned is the same as the vanishing of the commutator of $W(\tau)$ and $M$ (mathematically more precise: of $W(\tau)$ and $\one\otimes M$): 
\beq
[W(\tau),\;\one\otimes M]_{-}\,=\,
\big[U_{\mm}(\tau)\big(P^{A}_{j}\otimes W_{\sfM}(0)\big)
U^{\dagger}_{\mm}(\tau),\;\one\otimes M\big]_{-}=\zero\,,
\label{comm1MW}\enq
for \emph{every} pure inital state 
$P^{A}_{j}$ of system $\sfS$. 

The Lemma that Stein proves is, put slightly more abstractly,
as follows. Given bounded operators $M,Q\in\mathfrak{B}(\calH_{2})$, bounded operator $W\in\mathfrak{B}(\calH_{1}\otimes\calH_{2})$, and projector $P$ on $\calH_{1}$. If $\,P\otimes Q\,$ and $W$ commute, then there is a unique bounded operator $T_{Q}$ on $\calH_{2}$ such that the product of $P\otimes Q$ and $W$ can be written as $\otimes$-factorised operator $P\otimes T_{Q}$. As Stein (1997, p.$\,$236) points out in supplementary remark~3, a consequence of the Lemma is that the existence of $T_{Q}$ is sufficient and necessary for the commutativity of $\,P\otimes Q\,$ and $W$:
\beq
[P\otimes Q,\;W]\,=\,\textit{0}\;\;\Longleftrightarrow\;\;
\uniek\,T_{Q}\in\mathfrak{B}(\calH_{2}):\;
(P\otimes Q)W\,=\,P\otimes T_{Q}\,,
\label{SteinLemma}\enq
where $T_{Q}$ depends on $Q$ (whence the subscript) but \emph{does not depend} on~$P$.

The Insolubility Theorem is ``an immediate and sweeping consequence'' of this Lemma, as Stein 
(1997, p.$\,$237) puts it. 
Let the initial state of $\,\sfS\sqcup\sfM\,$ be
$\,W(0)=P^{A}_{j}\otimes W_{\sfM}(0)\,$ ---
system $\sfS$ is assumed to be in pure state
$P^{A}_{j}$ initially. The expectation-value for $M$
at the end of the measurement,
at time $\,t=\tau\,$, is by the trace-formula~\eqref{TrWP}:
\beq
\la 1\otimes M\ra_{W(\tau)}\,=\,\trm{Tr}\left(U_{\mm}(\tau)\big(P^{A}_{j}\otimes W_{\sfM}(0)\big)U^{\dagger}_{\mm}(\tau)\,(\one\otimes M)\right)\,.
\label{expM2}
\enq
The trace is invariant under cyclic permutation:
\beq
\la 1\otimes M\ra_{W(\tau)}\,=\,\trm{Tr}\left(\big(P^{A}_{j}\otimes W_{\sfM}(0)\big)U^{\dagger}_{\mm}(\tau)\,(\one\otimes M)U_{\mm}(\tau)\right)\,.
\label{expM3}
\enq

Since in general, operator $X$ commutes with 
$\,UYU^{\dagger}\,$
iff $Y$ commutes with $\,U^{\dagger}XU\,$,
it follows from \eqref{comm1MW} that
also $\,P^{A}_{j}\otimes W_{\sfM}(0)\,$ 
and $\,U^{\dagger}_{\mm}(\tau)\,(\one\otimes M)U_{\mm}(\tau)\,$ commute. According to the Lemma \eqref{SteinLemma}, when choosing
$P^{A}_{j}$ for $P$, $W_{\sfM}(0)$ for $Q$,
and $W(\tau)$ for $W$ ~\eqref{WSMtau},
there is a unique bounded operator $T\equiv T_{W_{\sfM}(0)}$ on $\calH_{\sfM}$,
which does not depend on $P^{A}_{j}$, and which is such that from \eqref{expM3} we obtain:
\beq
\la 1\otimes M\ra_{W(\tau)}\,=\,\trm{Tr}\big(P^{A}_{j}\otimes T\big)\,=\,\trm{Tr}\big(P^{A}_{j}\big)\trm{Tr}(T)\,=\,
\trm{Tr}(T)\,.\label{expM4}
\enq
Hence the expectation-value of the measurement
operator $M$ at time $t=\tau$
is \emph{independent} of the initial
state $P^{A}_{j}$ of system $\sfS$. In other words,
the probability measure over measurement outcomes of $A$ as determined by the initial state of $\sfS$ is not reproduced by the probability measure over values of the measurement operator $M$ of $\sfM$, thereby scandalising the Propability Reproducibility Condition. We have arrived at the
Probability Problem of Measurement Outcomes (p.$\,$\pageref{PPMO}).

\tbf{Moral.} The moral of the Insolubility Theorem is that  describing measurements unitarily (Universal Dynamics Postulate) in terms of mixed states, obeying the Propability Reproducibility Condition, breeds contradictions. (The absence of the Spectrum Postulate, the Eigenlink and SMOP among the premises leading to a contradiction we have already duly noted.) Some will draw the further moral that the Projection Postulate is inevitable. Standard and Copenhagen QM are off the hook. All interpretations of QM that reject the Projection Postulate and adopt a Universal Dynamics Postulate must face the Insolubility Theorem, which includes Modal Interpretations, Rovelli's Relational Interpretation, and of course Everett and Many Worlds. Whereas rejecting SMOP makes Everett and Many Worlds escape the Reality Problem of Measurement Outcomes, this option is unavailable in the face of the Probability Problem of Measurement Outcomes --- it may aggravate their `probability problem'. The only way to go, then, for adherents of the Universal Dynamics Postulate (only unitary measurements) seems to deny that measurement  devices initially never are in mixed states, but always in the pure ready-to-measure state when the measurement begins. Then one is safe.
\section{{\punt}The Reality Problem of the Classical World}
\label{SectRPCW}
The Reality Problem of Measurement Outcomes is a 
reality problem of properties of measurement devices
when described unitarily, granted the Eigenlink and
SMOP. We have seen how narrowly standard QM escapes the 
lethal inconsistency, due to its conditional Dynamics 
Postulate and its Projection Postulate. But there is another 
`reality problem' about properties not restricted to 
measurement devices but about \emph{all} actual physical systems that are not subjected to measurement, which is the overwhelming majority of physical systems in the universe --- \emph{nearly all} of them.

When we talk about the world that surrounds us, the world we  see, hear, smell, touch and feel, the \emph{observed observable world}, the `manifest world' (W.F.~Sellars), we do this mostly in terms of spatially extended material objects that have properties and are interrelated, subjects and their capacities included. These properties and relations may or may not change due to the influence that objects exert on each other (by means of physical interactions). But at each moment in time, every object and every subject possess several properties and exhibits several relations to other subjects and objects. This is often called the \emph{Classical World}, because the metaphysical picture just sketched, in terms of objects, subjects, possessed properties and exhibited relations, fits \emph{classical} physical theories like a  glove, as it does in fact all other scientific theories as well.\fn{Physical theories accepted from the Scientific Revolution in the 17th century onwards until 1900 have been baptized `classical'; the ones from 1900 onwards are then called `modern'.} Predicate Logic follows suit to cannonize the Classical World logically.

In terms of QM, the states of every two physical systems that have interacted in the past, or are interacting in the present, will generically be superpositions due to the Dynamics Postulate in bases of eigenstates that we associate with properties that we observe: the state of their composite systems is \emph{entangled}. But then the Eigenlink prohibits the attribution of these properties to the physical systems, and their interrelations when taken as properties of   composite systems. We have arrived at the following profound metaphysical problem.
\display{\donkrood{\ding{115}}~\tbf{IV. The Reality 
Problem of the Classical World.} 
\emph{How is the Classical World, a world
with physical systems possessing properties and
exhibiting relations, compatible with 
QM, specifically in the light of its Eigenlink and the generically entangled states of physical systems?}
\label{RPCW}}

The Reality Problem of Measurement Problems (p.$\,$\pageref{RPMO}) can be seen as a very special case of the Reality Problem of the Classical World, where we consider \emph{two} physical systems, a measurement device and a measured physical system, and let them interact unitarily, so that by vice of the Eigenlink we end up with two systems devoid of the \emph{properties} we believe they must have when the measurement has ended. To repeat, the Projection Postulate saves the day for standard QM. But this leaves physical systems in the universe \emph{unmeasured} by us without any properties and relations at all, which is, to repeat, nearly everything in the universe. The Classical World is lost. Interpretations of QM aim to regain the Classical World: it is the very reason of their existence.

A different manner to express \emph{roughly} the same problem are so-called \emph{problems of the classical limit}: when processes become slow and physical systems macroscopic (constituted by very many particles), what QM then says about them must be approximately (`in the limit') the same as what the appropriate classical physical theories say about them, e.g.\ classical mechanics, classical electro-dynamics, thermo-dynamics, optics. This are \emph{inter-theoretical} problems, about `limiting-relations' between QM and other physical theories, predicated on the assumption that these other physical theories describe the macroworld correctly. These problems of the classical limit, and the reverse problem, of `quantisation' (how to get from these theories to QM), has been and is an intense area of theoretical and mathematical research.\fn{E.g.\ Ehrenfest (19270, Landsman (1998), Bracken (2003).} Yet even if all the problems of classical limits have been solved, in the regimes where the mentioned theories fail and QM must take over, then in the world of the brief and the tiny, we still have neither properties nor relations. The Reality Problem of the Classical World then is at best partly solved, not completely.
\section{{\punt}The Measurement Explanation Problem}
\label{SectMIP}
The two postulates of standard QM that mutually exclude and jointly exhaust the change of state over time (Dynamics and Projection Postulate) evoke the question: why \emph{two}, and why \emph{these} two? More specifically, to measure something is to act, and to act is to do something with a purpose: gathering knowledge about a physical system in the case of measuring. To act is a manifestation of human agency. Indeed,  \emph{human} agency, because pickels, protons, peanuts, pandas and planets do not and cannot \emph{measure} anything. Measurement is an \emph{anthropomorphic} concept, and this concept occurs in both postulates governing the change of state of physical systems everywhere in the uninverse. This is without precedent in the history of physics, and perhaps of natural science. Von~Neumann spoke of two types of  processes in the universe, insipidly calling them \emph{Prozess~1} (measurement processes) and \emph{Prozess~2} (unitary processes, see Figure~1~(b), p.$\,$\pageref{Fig2}).  Measurement processes are indeterministic, discontinuous and non-linear, whereas unitary processes are determnistic, continuous and linear. Somewhat anachronistically, one could submit that Aristotle's distinction between artifical and natural processes has been resurrected, like Larazus from the dead. A why-question is a request for an explanation, so here we go:
\display{\donkrood{\ding{115}}~\tbf{V.~~The Measurement Explanation Problem.} Why is there an anthropomorphic concept of human agency, the concept of measurement, present in the postulates of a theory of inanimate matter (QM)? Why do physical interactions between physical systems obey anthropomorphic laws of nature as we use them in measurements? \label{MExplanP}}\par
\begin{center}
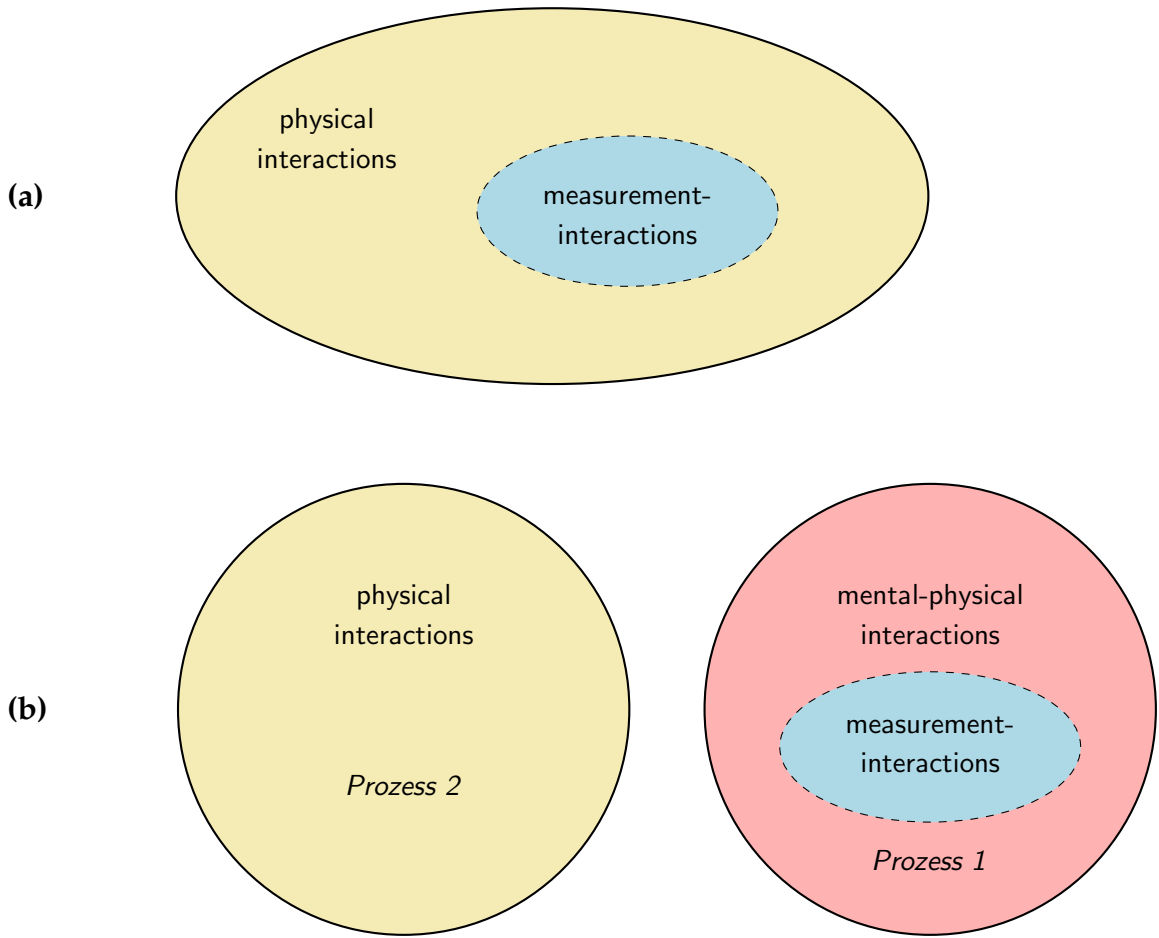
\begin{figure}[H]
\protect\vspace*{2em}
\begin{tikzpicture}
\draw[thick,fill=LightGoldenrod!60] (0,0) ellipse (5 and 2.5); 
\draw[fill=LightBlue,dashed] (1,-0.2) ellipse (2 and 1); 
\draw (1,0) node{\tsf{\small measurement-}};
\draw (1,-0.5) node{\tsf{\small interactions}};
\draw (-3,1) node{\tsf{\small physical}};
\draw (-3,0.5) node{\tsf{\small interactions}};
\draw (-7,0) node{\tbf{(a)}};
\end{tikzpicture}
\par\protect\vspace*{3em}
\begin{tikzpicture}
\draw[thick,fill=LightGoldenrod!60] (-3,0) circle (3); 
\draw[thick,fill=red!30] (4,0) circle (3); 
\draw (4,1.5) node{\tsf{\small mental-physical}};
\draw (4,1) node{\tsf{\small interactions}};
\draw (4,-2) node{\tsf{\small \emph{Prozess~1}}};
\draw (-3,1.5) node{\tsf{\small physical}};
\draw (-3,1) node{\tsf{\small interactions}};
\draw (-3,-1) node{\tsf{\small \emph{Prozess~2}}};
\draw[fill=LightBlue,dashed] (4,-0.5) ellipse (2 and 1); 
\draw (4,-0.2) node{\tsf{\small measurement-}};
\draw (4,-0.7) node{\tsf{\small interactions}};
\draw (-8,0) node{\tbf{(b)}};
\end{tikzpicture}
\par\protect\vspace*{1em}
\caption{\small \tbf{(a)} How today nearly everybody sees the relation between physical and measurement interactions:
measurement interactions are physical interactions. ~\tbf{(b)} How Von~Neumann and Wigner saw measurement interactions:
mental-physical interactions (consciousness causing collapse).}
\label{Fig2}
\end{figure}
\end{center}
\par \vspace*{-1em} Of course, we have already four distinct fundamental physical interactions that obey distinct laws: electro-magnetic, nuclear (`strong'), radio-active (`weak'), and grativational. Electro-magnetic interaction is a unification of electric and magnetic 
interaction. The Standard Model sort of 
unifies the electro-magnetic and the radio-active interaction in the electro-weak interaction, and hypothesises that in the very early universe,
the electro-weak and the nuclear force once were unified.
The Standard Model is unification on crutches.
The inclusion of gravity has become a head-ache dossier of theoretical physics: the Holy Grail of a theory of 
quantum gravity.
So what's the problem that we have a fifth type of interaction, the measurement interaction, governed by
yet another distinct law of nature? A hand full of interactions, with one distinctively human finger. What's the problem with that?

Well, for starters, nearly all measurement interactions 
\emph{are} electro-magnetic: this is simply how
measurement devices work, `mechanical' ones notably
included, as a moment of thought will reveal.
As soon as \emph{we} baptise an electro-magnetic 
interaction between physical systems a \emph{measurement interaction}, as soon as human agency is involved, it starts to obey a different law of nature. Why is that? Why does Mother Nature switch laws about the very same physical interaction as soon as \emph{we} show our faces?

Standard QM and 
Copenhagen QM face the Measurement Explanation Problem.
Most interpretations adopt a Universal Dynamics Postulate
and attempt to describe measurement interactions unitarily,
and then face some of the reality 
problems expositioned above, but
they do not face the Measurement Explanation
Problem (Figure~1~(a)). A related but distinct problem is the final measurement problem, to which we turn next.
\section{{\punt}The Measurement Meaning Problem}
\label{SectMEP}
Measurement, quantitative observation, qualifies as a species  of knowledge acquisition. And all men desire to know, as Aristotle wrote in the opening sentence of his \emph{Metaphysics}. To measure is, as mentioned in the previous Section, a manifestation of human agency, a species of behaviour: moving the human body, or parts of the human body, with a purpose. To measure physical magnitudes teaches physicists what values these magnitudes have, if only at the point of measurement. In all non-quantum physical theories, and in all theories in other scientific praxes, to measure is to reveal what determinate property the measured object possesses. Not so in standard QM. The ascription of determinate properties to measured physical systems happens  only at the time when the measurement result comes into being. The appearance of these determinate properties in the world seems to be an event of \emph{creatio ex nihilo}, as if experimenters and observers are Wizards performing acts of Metaphysical Magic. When we put it in metaphysical vocabulary as transforming a \emph{determinable} property into a \emph{determinate} property, we are at the level of being (\emph{ordo essendi}) rather than at the level of knowing (\emph{ordo cognoscenti}). The physics laboratory has become a place of Ontic Sorcery, rather than mere Epistemic Agency.

Well, let's not get carried away.
Perhaps better to say prosaically that the 
determinate properties are `produced
by' the measurement interaction between measured physical
system and measuring device. 

Let's state the next and last measurement problem:
\display{\donkrood{\ding{115}}~\tbf{VI. The Measurement Meaning Problem.} \emph{What is a measurement? What makes an interaction between physical systems a measurement
interaction? What makes a physical system a measurement
device?}\label{MExplicP}}

One famous and often-quoted piece of ranting and raving 
about this problem is Bell's (1990), his paper
`Against Measurement!' in \emph{Physics World}:
\display{\small What exactly qualifies some physical 
systems to play the role of `measurer'? Was the 
wavefunction of the world waiting to jump for thousands of 
millions of years until a single-celled living creature 
appeared? Or did it have to wait a little longer, for some 
better qualified system $\ldots$ with a PhD? If the theory 
is to apply to anything but highly idealised laboratory 
operations, are we not obliged to admit that more or less 
`measurement-like' processes are going on more or less all 
the time, more or less everywhere? Do we not have jumping 
then all the time? $(\ldots)$ The first charge against `measurement', in the fundamental axioms of quantum mechanics, is that it anchors there the shifty split of the world into `system' and `apparatus'. A second charge is that the word comes loaded with meaning from everyday life, meaning which is entirely inappropriateinthe quantum context. When it is said that something is `measured' it is difficult not to think of the result as referring to some pre-existing property of the object in question.}

Bell (1990) looks in vain in classic texts expounding standard 
QM (Dirac, Von~Neumann, Landau \& Lifshitz, Gottfried)
for clarity and rigour about what a measurement is.
Quantum-mechanical measurement theory (absent in the works mentioned by Bell) provides more detailed mathematical
representations of measurement interactions, but it leaves the concept of measurement, remarkably, un-analysed.
Take the authoritive monograph of Bush \emph{et al.}~(1996). In their introductory Section on `The Notion of Measurement', they write (\emph{ibid.}, p.$\,$25, their emphasis, our symbols):
\display{\small The purpose of measurements is the determination of properties of the physical system under investigation. In this sense the general conception of measurement is that of an unambiguous comparison: the object system $\sfS$, \emph{prepared} in a state $W$, is brought into a suitable contact --- a \emph{measurement coupling} --- with another, independently prepared system, the \emph{measuring apparatus} from which the \emph{result} related to the measured observable $A$ is \emph{determined} by \emph{reading} the value of the \emph{pointer observable} $M$. It is the goal of the quantum theory of measurement to investigate whether measuring processes, being physical processes, are the subject of quantum mechanics. This question, ultimately, is the question of the universality of quantum mechanics.}
The concept of a measurement is not analysed but taken
for granted.

In general, in philosophy, when faced with the problem of analysing a concept, we can walk two ways: Wittgenstein's Way and Carnap's Way. Let's take a walk.

\emph{Wittgenstein's Way.} In the opening page of \emph{The Blue Book}, Wittgenstein (1958, p.$\,$1), advances that to answer the question `What is length?', it helps to answer the question `How do we measure length?', and draws the analogy that to answer the 
question `What is the meaning of a word?', it is better
to ask: `How is this word used?' To ask what it means 
\emph{to measure} the momentum of a scattered elementary
particle in CERN is answered by an experimenter working
in CERN explaining you how they do it. To ask what
it means \emph{to measure} the temperature of gas in 
a vessel, an enigineer will show you a manometer
and will explain how this instrument works. To ask
what it means \emph{to measure} an electric current
in a circuit is answered by explaining how an ammeter
works, which is made part of the circuit. And so forth.
In general, what it means \emph{to measure} physical magnitude
$A$ of some physical system can
be explained by some relevant expert. Residues of unclarity will be 
cleared up by the expert whenever asked. When we have such 
explanations of every type of measurement performed by 
all relevant experts, then we are done.
What more \emph{is} there to explain? According to a 
use-conception of meaning, there is nothing more to explain. We may draw up lists of rules governing the use of the words `to measure' and `measurement'. Is the unsatisfied philosopher
not falling victim to the philosopher's craving for generality and essence?

We might seek something that all kinds of measurements
have in common, which could then characterise what the concept of measurement \emph{is}. This is presumably what Bell has been looking for, in vain. Bell craved for 
generality and essence, like a true philosopher. If there isn't something that all types of
measurement have in common, but every type of measurement
has something in common with some other types, then
measurement is what Wittgenstein baptised a 
\emph{family-resemblance concept}. If satisfied 
with such a conclusion, the Measurement Meaning Problem
evaporates, because it presupposes that all kinds of measurement
in physics have something in common, which must be 
captured by an explication. Bell and most philosophers
(of physics) will judge that to end the inquiry into measurement with this Wittgensteinian conclusion is a cop out. Wittgenstein's Way is not most philosophers' favourite way. 
They prefer Carnap's Way.

\emph{Carnap's Way.} 
An \emph{explication} of a concept is a criterion for 
that concept, which is a condition that is both sufficient 
and necessary. An explication must be an explicit
logical combination of other concepts, and should besides
being extensionally correct 
also be intensionally correct.\fn{And must meet a few other conditions we gloss over. See Carnap (1950).} Intensional correctness is that 
\emph{explicans} and \emph{explicandum}
must be synonymous. Extensional correctness is that the same
things fall under the extension of both 
\emph{explicans} and \emph{explicandum}. Inspection
of how the concept of measurement 
\emph{is used} when walking Carnap's Way
is as unavoidable as it is when walking Wittgenstein's 
Way. In a \emph{Liber Amicorum}
for P.C.\ Suppes, yours truly took a stab at finding an
explication of the concept of measurement;
we end this Section by summarising this 
explicaton, with slight improvements.\fn{See Muller [2015] for elucidation of the various features of this explication.}

We begin by recalling that a physical system $\sfS$ is 
\emph{observable} (to us, human beings) ~iff~ whenever an arbitrary healthy human being were in front of $\sfS$ in broad daylight, and were looking at $\sfS$, she would see $\sfS$.\fn{Cf.\ Muller (2005), Votsis (2015).} Next observation predicates.
\display{\tbf{Criterion for an Observation Predicate.} 
A predicate $F$ applied to observable physical system $\sfS$ is an \emph{observation predicate} ~iff~ whenever
an arbitrary healthy human being were in front of $\sfS$ 
in broad daylight, and were looking at $\sfS$, then she either would immediately judge that $F(\sfS)$, or 
judge that $\,\neg F(\sfS)$, relying only on 
looking at~$\sfS$, not making any inferences or 
appealing to some theory. (Rather than in terms of judgement, one can phrase this criterion also in terms of immediately
obtaining an occurrent perceptual belief.)}
Next a criterion for physical system $\sfS$
being a piece of measurement apparatus.
\display{\tbf{Criterion for an $A$-Measurement Apparatus.}~~Physical system $\sfM$ is 
an $A$-\emph{measurement apparatus} ~iff\\
(M1)~~$\sfM$ is observable;\\
(M2)~~there is a correlation between observation predicates $F$ of the type `$\sfM$ displays value $a$',
and sets of values of $A$; and \\
(M3)~~the correlation of (M2) is the result of the 
$A$-relevant physical 
interaction between $\sfM$ and physical system $\sfS$, of which $A$ is a determinable.}
Friends of causality can replace `is the result of' in (M3) with: is caused by. A physical interaction between $\sfS$ and $\sfM$ is $A$-\emph{relevant} iff the interaction is needed to explain why the correspondence in (M2) obtains. For friends of causality, this explanation will then be a causal explanation. 

The explanation of the Ontic Sorcery of determinable physical properties becoming determinate at the end of a measurement (granted the Projection Postulate) ought to be part of the explanation mentioned in the criterion of an $A$-relevant measurement~(M3). Since Modal Interpeters take $A$-measurements to reveal possessed determinate properties  $\la A,a_{j}\ra$, they will prefer a different explanation ---  there is no Ontic Sorcery going on in laboratories according to Modal Interpreters. We see that the explication of what an $A$-measurement apparatus is has a feature that depends on which interpretation of QM is at play; but only there, within the explanation in (M3) of the correlation in~(M2).

Finally, an explication of what it means to measure something
by a human being (or by any other being in the universe that
has comparable capacities):
\display{\tbf{Criterion for Measurement.}~~Human beings \emph{measure} physical magnitude $A$ of physical system $\sfS$ by means of $A$-measurement apparatus $\sfM$ and obtain 
value $a$ ~iff~ they make $\sfS$ and $\sfM$ physically interact $A$-relevantly, and this $A$-relevant interaction
results in ascribing value $a$ to $A$, 
which value $\sfM$ registers or displays.}

\begin{figure}[H]
\begin{tikzpicture}[>=stealth,scale=0.90]
\draw[fill, color=DarkGreen!15] (-6,10) rectangle (-2,11);
\draw[very thick, color=DarkGreen] (-6,10) rectangle (-2,11);
\draw (-4,10.5) node{\tsf{\small human vision}}; 
\draw[->,very thick,color=DarkGreen] (-4,9.9)--(-1,7.1);
\draw[fill, color=blue!15] (6,10) rectangle (2,11);
\draw[very thick, color=blue] (6,10) rectangle (2,11);
\draw (4,10.5) node{\tsf{\small light}};
\draw[->,very thick,color=blue] (4,9.9)--(1,7.1); 
\draw[fill, color=DarkGreen!15] (-2,6) rectangle (0,7);
\draw[fill, color=blue!15] (0,6) rectangle (2,7);
\draw[very thick, color=blue!50!DarkGreen] (-2,6) rectangle (2,7);
\draw (0,6.5) node{\tsf{\small observability}}; 
\draw[->,very thick,color=blue!50!DarkGreen] (-1,5.9)--(-6,3.1);
\draw[->,very thick,color=blue!50!DarkGreen] (0,5.9)--(0,0.1);
\draw[fill, color=DarkGreen!15] (-9,6) rectangle (-5,7);
\draw[very thick, color=DarkGreen] (-9,6) rectangle (-5,7);
\draw (-7,6.5) node{\tsf{\small belief/judgement}}; 
\draw[->,very thick,color=DarkGreen] (-7,5.9)--(-7.5,3.1);
\draw[fill, color=Chocolate4!20] (9,6) rectangle (5,7);
\draw[very thick, color=Chocolate4] (9,6) rectangle (5,7);
\draw (7,6.5) node{\tsf{\small explanation}};
\draw[->,very thick,color=Chocolate4] (7,5.9)--(6.5,3.1);
\draw[fill, color=DarkGreen!15] (-9,2) rectangle (-6.5,3);
\draw[fill, color=blue!15] (-6.5,2) rectangle (-5,3);
\draw[very thick, color=blue!50!DarkGreen] (-9,2) rectangle (-5,3);
\draw (-7,2.5) node{\tsf{\small observation predicate}};
\draw[->,very thick,color=blue!50!DarkGreen] (-7,1.9)--(-1,0.1);
\draw[fill, color=blue!15] (9.5,2) rectangle (7,3);
\draw[fill, color=Chocolate4!20] (7,2) rectangle (4.5,3);
\draw[very thick, color=blue!40!Chocolate4] (9.5,2) rectangle (4.5,3);
\draw (7,2.5) node{\tsf{\small $A$-relevant interaction}};
\draw[->,very thick,color=blue!40!Chocolate4] (6,1.9)--(1,0.1);
\draw[->,very thick,color=blue!40!Chocolate4] (7.5,1.9)--(5,-2.9);
\draw[fill, color=DarkGreen!15] (-2.5,-1) rectangle (-1,0);
\draw[fill, color=blue!15] (-1,-1) rectangle (1,0);
\draw[fill, color=Chocolate4!20] (1,-1) rectangle (2.5,0);
\draw[very thick, color=blue!40!Chocolate4] (-2.5,-1) rectangle (2.5,0);
\draw (0,-0.5) node{\tsf{\small $A$-measurement apparatus}}; 
\draw[->,very thick,color=blue!40!Chocolate4] (0,-1.1)--(4.5,-2.9);
\draw[fill, color=Chocolate4!20] (7,-4) rectangle (5.5,-3);
\draw[fill, color=blue!15] (5.5,-4) rectangle (4,-3);
\draw[fill, color=DarkGreen!15] (4,-4) rectangle (3,-3);
\draw[very thick] (7,-4) rectangle (3,-3);
\draw (5,-3.5) node{\tsf{\small $A$-measurement}}; 
\draw (0,-4.5) node{\textcolor{white}{nothing}};
\end{tikzpicture}
\caption{\small Conceptual dependencies of the concept of measurement as explicated in the current paper, starting
with the concepts of human vision, light, belief (or judgement), and explanation. \label{Fig1}}
\end{figure}
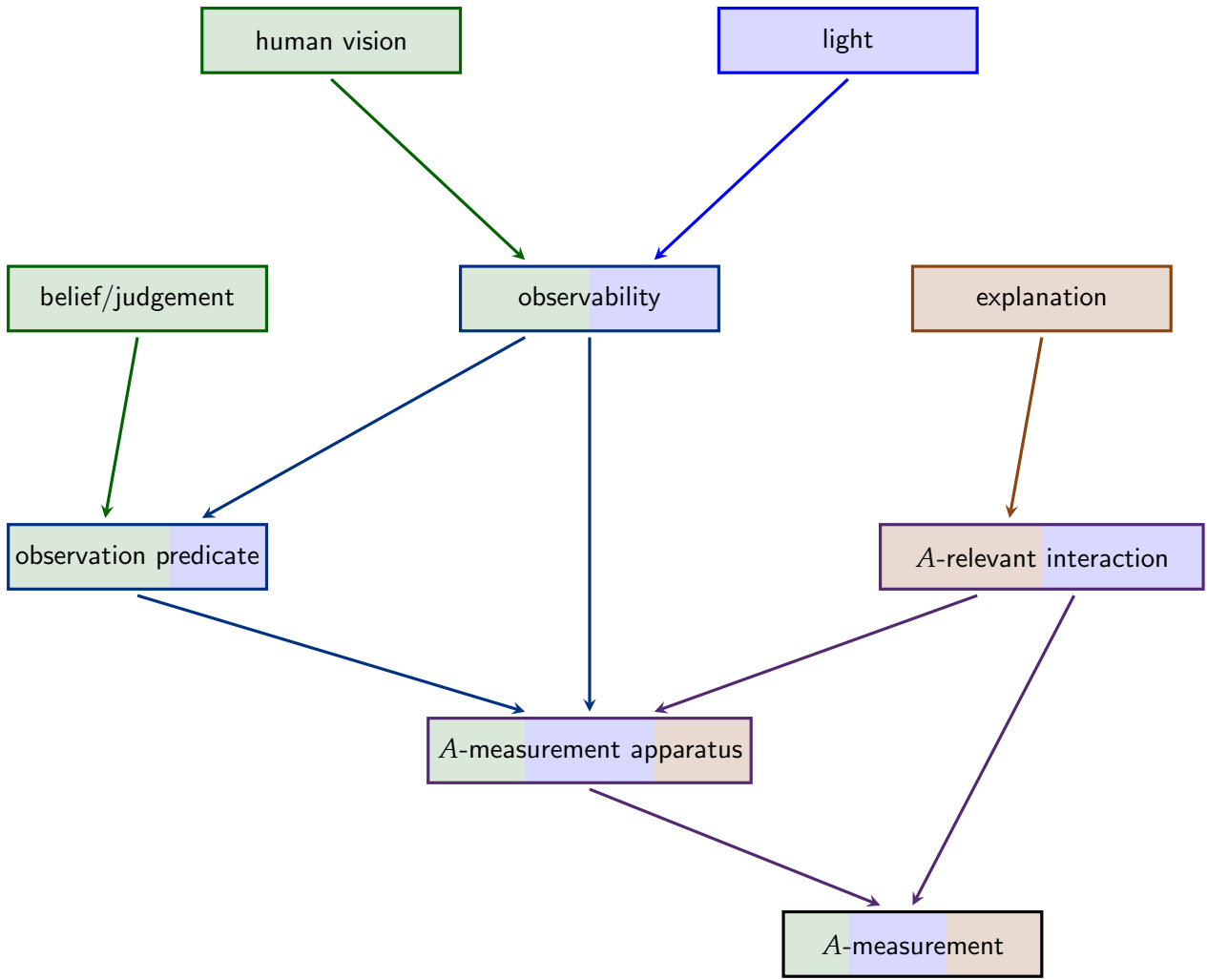

The conceptual dependencies are depicted in 
Figure~\ref{Fig1}~(p.$\,$\pageref{Fig1}). 

Some measurements in physics, e.g.\ time measurements, by means of clocks, do not seem to meet the criterion above --- with what physical system does a clock interact? Yet the criterion does fit measurement theory of QM seemlessly, as expounded in e.g.\ Bush \emph{et al.}~(1996). Further, every interpretation of QM could adopt this explication of the concept of measurement; it is interpretation neutral, or so we claim.
\section{{\punt}Recapitulation}
\label{SectRecap}
We have distinguished six distinct `measurement problems'
about QM, which are not all problems that 
\emph{standard} QM must face.
Three problems are \emph{polylemma problems}:
they present three bouquets of inconsistent
propositions, and force one to choose which
proposition of each bouquet to renounce.
One problem is an \emph{how-to problem}, 
and another is a \emph{why-problem} and 
therefore is a request for an explanation. 
The sixth problem is a \emph{what-problem}:
a request for an 
explication, of the concept of measurement.

The \emph{first} problem is the Reality Problem of 
Measurement Outcomes (p.$\,$\pageref{RPMO}), which is a logical clash between three 
plausible 
propositions, granted the State, Magnitude and Spectrum 
Postulate (of standard QM): that all physical interactions 
are unitary (Universal Dynamics Postulate), that physical 
systems have properties iff their state is in the relevant 
eigenstate (Eigenlink), and that properly functioning 
measurement devices yield a single outcome upon measurement 
(SMOP). Standard QM escapes the contradiction by restricting
unitary evolution to when no measurements are performed,
and adopts the Projection Postulate for when measurements
are performed. 

The \emph{second} problem is the State 
Completess Problem (p.$\,$\pageref{StateCompl}), 
which states that standard QM is 
committed to the measurement- as well as the 
property-incompleteness of the state. 
Friends of standard QM take this to express the indeterministic character of QM, and of 
microphysical reality. Not really a \emph{problem},
unless one is a determined determinist. Then one is
in trouble, big time.

The \emph{third} problem is the Probability Problem of Measurement Outcomes, which states, granted only the Mixed State and Magnitude Postulate: the Probability Postulate is incompatible with all physical interactions being unitary and obeying the Property Revealing Conditon.

Whereas the Reality Problem of Measurement Outcomes is about determinate properties and employs the Eigenlink and SMOP (but does not employ the Probability Postulate), the Probability Problem of Measurement Outcomes (p.$\,$\pageref{PPMO}) is about measurement outcomes and does  employ neither the Eigenlink nor SMOP (but does employ the Probability Postulate). Both these problems create enormous problems for taking measurement interactions to be unitary. The first problem uses only the Property Revealing Condition (measurements reveal possessed properties) to deduce a contradiction, the second problem the Probability Reproductibility Condition (the probability distribution of measurement outcomes must reflect the initial probability distribution of the physical magnitude measured). Both conditions are necessary for unitary evolutions to qualify as measurement interactions, and they are jointly sufficient:
\beq
t\mapsto U(t)~~\trm{measurement~~~iff~~~Prop.\ Rev.\ Cond. ~and~ Prob.\ Reprod.\ Cond.}
\enq

Standard QM, which rejects measurement interactions to
be unitary and thereby escapes the first two inconsistency problems, does not escape the \emph{fourth} problem,
the Reality Problem of the Classical World (p.$\,$\pageref{RPCW}), of \emph{how to reconcile} 
the fact that physical systems we observe all around us
with their properties generically are not in eigenstates. 
That such a general metaphysical conceptual framework
could possibily clash with a scientific theory was
inconceivable before the advent of QM. Yet this is how
deep QM drills metaphysically. 

The \emph{fifth} problem is a request for an explanation
to friends of standard QM, when the Projection Postulate
has been adopted and the Dynamics Postulate has been 
restricted: the Measurement Explanation Problem 
(p.$\,$\pageref{MExplanP}). 
\emph{Why} does a manifestation of human agency occur in laws of nature governing all matter in the universe? 

The \emph{sixth} and final problem is a problem that has not been served with solutions over the past eighty years, say; it 
is a request for an explication of the concept of measurement
(the Measurement Meaning Problem, 
p.$\,$\pageref{MExplicP}), to answer the question
what measurement is. We summarised an attempted solution to this Meaning Problem.

D\'{e}cio Krause is a Brazilian philosopher who appreciates clarity, precision, and rigour eminently, who is fascinated by QM, and  who has payed attention to QM in several publications, notably  about indiscernibles, vague objects and quantum  logic.\fn{Krause (2000, 2007, 2010), French and Krause (1995, 2006), Krause and Da~Costa (1997).} In spite of the fact that my contribution does not relate directly to any of the issues in QM D\'{e}cio has addressed, I hope, and suspect that he will appreciate --- eminently or not --- the disentanglement of the six different problems that since the inception of QM have made, and are making, waves under the flag of `the measurement problem'. Arguably the Reality Problem of the Classical World is the flagship of these problems. But let's not forget that the flagship heads a small fleet. \\[2em]
{\small \emph{Acknowledgments.}\\
Thnx to Guido Bacciagaluppi, Maura Burke, Dennis Dieks, Menno Hellinga, Ronnie Hermens, Sam Rijken en Raoul Titulaer 
for comments, corrections, and suggestions.\\[1em]
\emph{Affiliations.} Erasmus School of Philosophy, Erasmus University Rotterdam,
Burg.\ Oudlaan 50, 3062 PA Rotterdam; and:
Descartes Centre for the History and Philosophy of Science,
Faculty of Science, Utrecht University, Utrecht; both in The Netherlands; E-mail: f.a.muller@uu.nl.}\\[3em]
\noindent\tbf{\large References}
\addcontentsline{toc}{section}{References}%
{\small \begin{refs}
{\item Bacciagaluppi, G., `Insolubility from No-Signalling',
\emph{International Journal for Theoretical Physics}
\tbf{53} (2014) 3465--3474.}
{\item Bell, J.S., `Against Measurement',
\emph{Physics World} \tbf{3} (1990), 33--40.}
{\item B\'{e}ziau, J.-Y.\ \emph{et al.}\ (eds.), 
\emph{Conceptual Clarifications.\ Tributes to
P.C.\ Suppes (1922--2014)}, College Press, 2015.}
{\item Bracken, A.J. (2003). `Quantum mechanics as an approximation to classical mechanics in Hilbert space', \emph{Journal of Physics A} \tbf{36.23} (2003): L329--L335.}
{\item Brown, H.R., `The Insolubity Proof of the Quantum
Measurement Problem', \emph{Foundations of Physics}
\tbf{16} (1986) 857--870.}
{\item Bub, J., \emph{The Interpretation of Quantum Mechanics}, Dordrecht and Boston: D.\ Reidel Publishing Company, 1975.}
{\item Bush, P., Shimony, A., `Insolubility of the Quantum Measurement Problem for Unsharp Observables', 
\emph{Studies in the History and Philosophy of Modern Physics} \tbf{27.4} (1997) 397--404.}
{\item Bush, P., \emph{et al.}, \emph{The Quantum Theory of Measurement} (2nd.\ Rev.\ Ed.), Berlin: Springer-Verlag, 1996}
{\item Carnap, R., \emph{Logical Foundations of Probability}, Chicago: University of Chicago Press, 1950.}
{\item Dirac, P.A.M., \emph{The Principles of Quantum Mechanics}, Cambridge: Cambridge University Press, 1928}
{\item Einstein, A., Podolsky, B.Y., Rosen, N. `Can the 
quantum-mechanical description of physical reality be considered complete?', \emph{Pysical Review} \tbf{35} (1935) 777--781.}
{\item Ehrenfest, P., `Bemerkung \"{u}ber die angen\"{a}herte G\"{u}ltigkeit der klassischen Mechanik innerhalb der Quantenmechanik', \emph{Zeitschrift f\"{u}r Physik} \tbf{45.7} (1927): 455--457.}
{\item Fine, A., \emph{The Shaky Game,\ Einstein, Realism and the Quantum Theory}, Chicago: University of Chicago Press, 1986.}
{\item Fine, A., `Insolubility of the Quantum Measurement
Problem', \emph{Physical Review D} \tbf{2} (1970) 2783--2787.}
{\item Krause, D., French, S., `Quantum sortal predicates', 
\emph{Synthese} volume \textbf{154} (2007) 417--430.} 
{\item French, S., Krause, D. (2006), \emph{Identity in Physics: A Historical, Philosophical and Formal Analysis}, Oxford: Clarendon Press.}
{\item French, S., Krause, D., `Vague Identity and 
Quantum Non-Individuality', \emph{Analysis}
\tbf{55.1} (1995) 20--26.}
{\item Krause, D., `Logical Aspects of Quantum (Non-)Individuality', \emph{Foundations of Science} \tbf{15} (2010) 79--94.}
{\item Krause, D., `Remarks On Quantum Ontology',
\emph{Synthese} volume \textbf{125} (2000) 155--167.}
{\item Krause, D., Da~Costa, N.C.A., `An Intensional 
Schr\"{o}dinger Logic', \emph{Notre Dame Journal of Formal Logic} \tbf{38.2} (1997) 179--194.} 
{\item Landsman, N.P., \emph{Foundations of Quantum Theory:
from Classical Concepts to Operator Algebras}, Berlin: Springer-Verlag, 2017.}
{\item Landsman, N.P., \emph{Topics between Classical Mechanics and Quantum Mechanics}, Berlin: Springer-Verlag, 1998.}
{\item Maudlin, T., `Three Measurement Problems',
\emph{Topoi} \tbf{14} (1995) 7--15.}
{\item Muller, F.A., `Circumveiloped by Obscuritads.\ The nature of interpretation in quantum mechanics: hermeneutic circles and physical reality, with cameos of James Joyce and  Jacques Derrida',
in: B\'{e}ziau \emph{et al.}\ [2015], pp.$\,$107--136.}
{\item Muller, F.A., `The Deep Black Sea: Observability and Modalify Afloat',
\emph{British Journal for the Philosophy of Science} \textbf{56} (20005), pp.~61--99.}
{\item Neumann, J.\ von., \emph{Mathematische Grundlagen der Quantenmechanik}, Berlin: Springer-Verlag, 1932.}
{\item Neumann, J.\ von, `Wahrscheinlichkeitstheoretischer Aufbau der Quantenmechanik', \emph{G\"{o}ttinger Nachrichten}
\tbf{1} (1927) 245--272.}
{\item Norsen, T., \emph{Foundations of Quantum Mechanics.\
An Exploration of the Physical Meaning of Quantum Theory},
Cham: Springer, 2017.}
{\item Rovelli, C., \emph{Helgoland: Making Sense of the Quantum Revolution}, London: Penguin, 2021.}
{\item Shimony, A., `Approximate Measurements in Quantum Mechanics II', \emph{Physical Review D} \tbf{9} (1974) 2321--2323.}
{\item Stein, H., `Maximal Extension of an Impossibility
Theorem concerning Quantum Measurement', in: \emph{Potentiality, Entanglement and Passion-at-a-Distance}, R.S.\ Cohen \emph{et al.}\ (eds.), Dordrecht: Kluwer Academic Publishers, 1997.}
{\item Votsis, I., `Perception and Observation Unladed',
\emph{Philosophical Studies} \tbf{172} (2015) 563--585.}
{\item Wigner, E.P., `The Problem of Measurement',
\emph{American Journal of Physics} \tbf{31} (1963) 6--15.}
{\item Wigner, E.P., `Remarks on the Mind-Body Problem',
in: \emph{The Scientist Speculates}, I.J.\ Good (ed.), London: Heinemann, 1961, pp.$\,$284--302.}
{\item Wittgenstein, L., \emph{The Blue and the Brown Book},
Oxford: Blackwell Publishing, 1958.}
\end{refs} }
\end{document}